\documentclass[twocolumn,floats,prl,aps,unsortedaddress,superscriptaddress]{revtex4-1}

\usepackage[dvips]{graphicx}
\usepackage{amsfonts}
\usepackage{amsmath}
\usepackage{amssymb}
\usepackage{exscale}
\usepackage{color}	%% FOR COLORED NOTES IN DRAFT. COMMENT THIS LINE FOR FINAL VERSION 
\usepackage{braket}	%% For nice BRAKET notation
\usepackage{multirow}	%% For columns spanning multiple rows in Tables
\usepackage{threeparttable} %%% For nice footnotes and references in tables
\usepackage{epstopdf}
\usepackage[normalem]{ulem}    %% To use \uline for underlining, 
\begin{document}
%%%%%%%
\title{Two-dimensional electron systems in ATiO$_3$ perovskites 
	   (A$=$~Ca,~Ba,~Sr): control of orbital hybridization and order}
%%%%%%%

\author{T.~C.~R\"odel}
\altaffiliation[Present address: ]{Laboratory for Photovoltaics, Physics and 
			Material Science Research Unit, University of Luxembourg, 
			L-4422 Belvaux, Luxembourg}
\affiliation{CSNSM, Univ. Paris-Sud, CNRS/IN2P3, Universit\'e Paris-Saclay, 
			91405 Orsay Cedex, France}
\affiliation{Synchrotron SOLEIL, L'Orme des Merisiers, Saint-Aubin-BP48, 
			91192 Gif-sur-Yvette, France}
\author{M.~Vivek}
\affiliation{Laboratoire de Physique des Solides, CNRS, Univ. Paris-Sud, 
			Universit\'e Paris-Saclay, 91405 Orsay Cedex, France}
\author{F.~Fortuna}
\affiliation{CSNSM, Univ. Paris-Sud, CNRS/IN2P3, Universit\'e Paris-Saclay, 
			91405 Orsay Cedex, France}
\author{P.~Le~F\`evre}
\affiliation{Synchrotron SOLEIL, L'Orme des Merisiers, Saint-Aubin-BP48, 
			91192 Gif-sur-Yvette, France}
\author{F.~Bertran}
\affiliation{Synchrotron SOLEIL, L'Orme des Merisiers, Saint-Aubin-BP48, 
			91192 Gif-sur-Yvette, France}
\author{R.~Weht}
\affiliation{Gerencia de Investigaci\'{o}n y Aplicaciones, 
			Comisi\'{o}n Nacional de Energ\'{i}a At\'{o}mica, 
			Avenida General Paz y Constituyentes, 1650 San Mart\'{i}n, Argentina}
\affiliation{Consejo Nacional de Investigaciones Cient\'{i}ficas y T\'ecnicas (CONICET), 
			Buenos Aires, Argentina}
% \affiliation{Instituto S\'{a}bato, Universidad Nacional de San Mart\'{i}n – CNEA, 
%			1650 San Mart\'{i}n, Argentina}
%%
% \author{M.~J.~Rozenberg}
% \affiliation{Laboratoire de Physique des Solides, CNRS, Univ. Paris-Sud, 
%			Universit\'e Paris-Saclay, 91405 Orsay Cedex, France}
%%
\author{J.~Goniakowski}
\affiliation{Institut des Nanosciences de Paris, UMR 7588, CNRS and Universit\'e Paris-6,
			4 place Jussieu, 75252 Paris cedex 05, France}
\author{M.~Gabay}
\affiliation{Laboratoire de Physique des Solides, CNRS, Univ. Paris-Sud, 
			Universit\'e Paris-Saclay, 91405 Orsay Cedex, France}
\author{A.~F.~Santander-Syro}
\email{andres.santander@csnsm.in2p3.fr}
\affiliation{CSNSM, Univ. Paris-Sud, CNRS/IN2P3, Universit\'e Paris-Saclay, 
			91405 Orsay Cedex, France}
%%%%%%%

%%%%%%%%%%%%%
\begin{abstract}
	%%%
	We report the existence of a two-dimensional electron system (2DES) 
	at the $(001)$ surface of CaTiO$_3$. 
	Using angle-resolved photoemission spectroscopy, we find a hybridization 
	between the $d_{xz}$ and $d_{yz}$ orbitals,
	% due to the rotation of the oxygen octahedra, 
	% which breaks the cubic symmetry of the ideal perovskite lattice. 
	% This hybridization is 
	not observed in the 2DESs at the surfaces 
	of other ATiO$_3$ perovskites, \textit{e.g.} SrTiO$_3$ or BaTiO$_3$. 
	Based on a comparison of the 2DES properties in these three materials, 
	we show how the electronic structure of the 2DES 
	(bandwidth, orbital order and electron density) 
	is coupled to different typical lattice distortions in perovskites.
	The orbital hybridization in orthorhombic CaTiO$_3$ results from the rotation
	of the oxygen octahedra, which can also occur at the interface of 
	oxide heterostructures to compensate strain. 
	%%
	% As rotations of oxygen octahedra are a common phenomena to compensate strain in oxides, 
	% the study of CaTiO$_3$ gives insights of effects occurring in 2DESs at the interface 
	% of oxide heterostructures showing lattice mismatch. 
	More generally, the control of the orbital order in 2DES by choosing different A-site cations 
	in perovskites offers a new gateway towards 2DESs in oxide heterostructures beyond SrTiO$_3$. 
	%%%
\end{abstract}
%%
%\pacs{73.20.-r, 73.20.At, 73.22.-f, 77.84.Ek}
\maketitle
%%%%%%

%%%%%%%%%%%%%%%%%%%%%%%%%
%%% INTRODUCTION
%%%%%%%%%%%%%%%%%%%%%%%%%
%%{\it Introduction.-} 
%%
%%%%%%%%%%%%
%% Figure 1
%%%%%%%%%%%%
\begin{figure*}[th]
  \begin{center}
   	  \includegraphics[clip, width=0.8\textwidth]{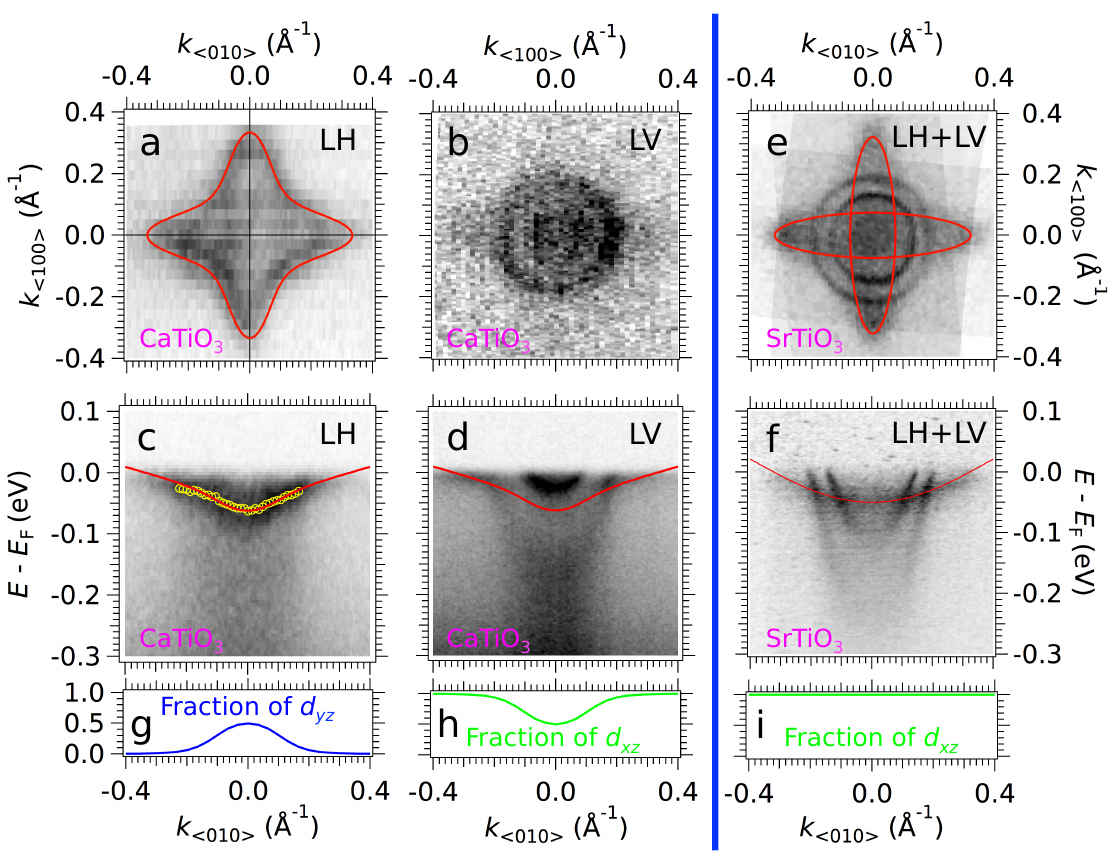}
  \end{center}
  \caption{\label{fig:CTO} 
      	%%%
  		(a,~b) Fermi surface intensity maps of the 2DES measured 
  		at the surface of CaTiO$_3$(001) 
  		close to $\Gamma_{005}~(h\nu=57$~eV) using linear horizontal (LH) polarization,
  		and close to $\Gamma_{115}~(h\nu=67$~eV) using linear vertical (LV)
  		polarization, respectively.
  		(c,~d) $E-k$ intensity maps measured at $\Gamma_{005}$ using LH and LV polarization. 
  		The red curves are based on a one-layer tight-binding model 
  		assuming orbital hybridization between the $d_{xz}$ and $d_{yz}$ orbitals.
  		The yellow markers in (c) are the peak positions of the fits 
  		of the energy distribution curves.
		(e,~f) Fermi surface and $E-k$ map corresponding to the electronic structure 
		of the 2DES at the (001) surface of SrTiO$_3$. 
		The shown intensity maps are a superposition of measurements 
		using LH polarization at $h\nu=90$~eV and LV polarization at $h\nu=47$~eV. 
		The red curves are, in this case, based on a tight-binding model 
		\emph{without} hybridization between the different $t_{2g}$ orbitals.
		(g,~h,~i) Momentum-resolved fraction of orbital character of the $d_{xz}$ 
		or $d_{yz}$ band visible in the $E-k$ maps in (c,~d,~f) 
		based on the tight-binding model showing the orbital hybridization 
		in the 2DES at the (001) surface of CaTiO$_3$. 
  		}
\end{figure*}
%%%%%%%%%%%%%%%%
%%
%%%%%%%%%%%%%%%%%
%% Figure DFT
%%%%%%%%%%%%%%%%%
\begin{figure}[th]
  \begin{center}
   	  \includegraphics[clip, width=0.49\textwidth]{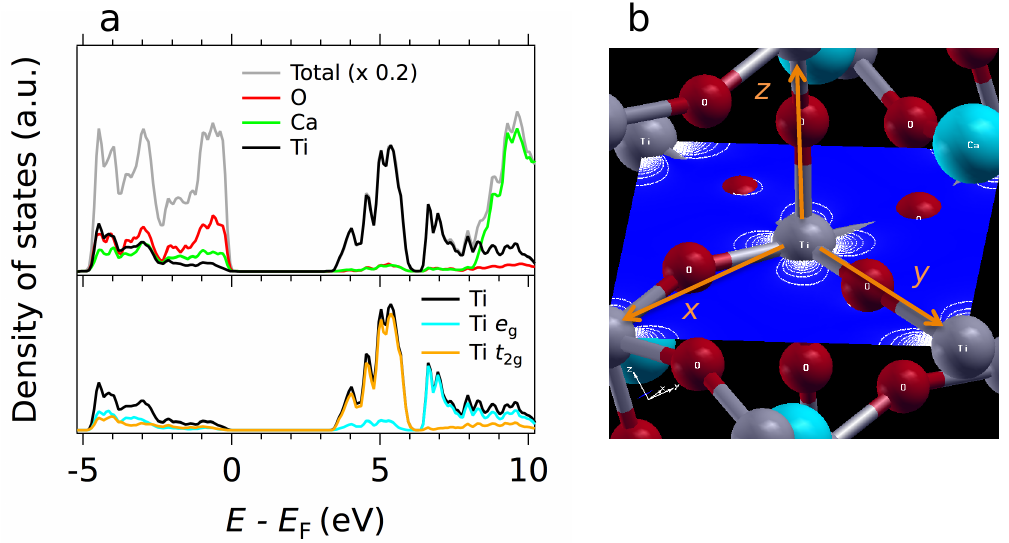}
  \end{center}
\caption{\label{figmain1} 
		%%%%
		(a) Top: Total, Ti-, Ca- and O-projected densities of states of bulk CaTiO$_3$ 
		obtained from HSE calculations. 
		% with soft oxygen pseudo-potential.
		%%
		Bottom: Decompositions into Ti $e_g$ ($d_{z^2} + d_{x^2 - y^2}$) 
		and Ti $t_{2g}$ ($d_{xy} + d_{yz} + d_{xz}$) components 
		shows that the conduction band minimum of CaTiO$_3$ 
		is mainly formed of $t_{2g}$ orbitals.
		%%%
		(b) Cut into the pseudo-TiO$_2$ plane (blue plane)
		of the charge density plot (white contours) for the lower energy state 
		of the conduction band in CaTiO$_3$. 
		Ca, Ti, and O atoms are represented by cyan, grey and red spheres, respectively.
		Orange arrows show the $(x, y, z)$ axes of the pseudo-cubic unit cell.
		%%%	
		}
\end{figure}
%%%%%%%%%%%%

%%%%%%%
{\it \bf Introduction.-} 
ABO$_3$ perovskites, where B is a transition-metal (TM) ion,
present many appealing phenomena, including ferroelectricty, ferromagnetism, superconductivity 
and strong electron-correlations~\cite{Dagotto2008, Zubko2011}. 
One reason for such a diversity is that the perovskite lattice can accommodate 
a large variety of differently sized A and B cations 
as described by Goldschmidt's tolerance factor~\cite{Goldschmidt1926}. 
This factor can be widely varied by the size of the A\text{--}site cation, 
resulting in different lattice distortions 
which strongly influence the electronic structure~\cite{Medarde1997, Kimura2003}.
%%%

Such a variety of functionalities within the same oxide family, 
together with the epitaxial compatibility amongst many of its members,
has boosted the interest in oxide heterostructures over the last two decades.
One prime example of these emerging properties 
is the two-dimensional electron system (2DES) found at the LaAlO$_3$/SrTiO$_3$ 
interface~\cite{Ohtomo2004}, which shows a wide range of properties including 
the coexistence of superconductivity and magnetism~\cite{Dikin2011, Li2011a} 
as well as a possibly unusual electron pairing mechanism~\cite{Cheng2015}. 
%%%%%
More recently, the discoveries of 2DESs at bare surfaces of various perovskites 
such as the paraelectric SrTiO$_3$~\cite{Santander-Syro2011,
Meevasana2011,Wang2014,Roedel2014,Walker2014}, 
the strong spin-orbit coupled KTaO$_3$~\cite{King2012,Santander-Syro2012,Bareille2014},
the catalyst TiO$_2$ anatase~\cite{Roedel2015}, or the ferroelectric BaTiO$_3$~\cite{Rodel2016}
triggered new avenues of research by providing deep insight 
into the microscopic electronic structure of such 2DESs, including orbital order, symmetries, 
and electron-phonon interaction effects~\cite{Santander-Syro2011, Meevasana2011, 
Chen2015e, Wang2015a}.
%%%

%%%%%%%%%%%%%%%%%%%%
%%% HERE WE SHOW %%%
%%%%%%%%%%%%%%%%%%%%
Here we report the discovery of a 2DES at the $(001)$ surface of CaTiO$_3$.
Moreover we find a significant hybridization between the $d_{xz}$ and $d_{yz}$ orbitals 
forming the 2DES, not observed in the 2DES's at the surface of other perovskite titanates, 
and show that it is induced by the rotation of the oxygen octahedra 
in the orthorhombic lattice resulting from the small size of the Ca ion. 
This is very appealing, as the possibility to use octahedral tilts to control the properties 
of oxide interfaces, such as magnetism, 
has attracted much attention lately~\cite{Rondinelli2012a,Ganguli2014,Liao2017}. 
To further explore the connection between lattice distortions in the perovskites 
and electronic structures in 2DES, 
we compare the 2DESs measured by ARPES 
at the surface of different titanates, ATiO$_3$ (A~$=$~Ca,~Ba,Sr)~\cite{Rodel2016}. 
We thus show that the orbital order, orbital symmetries 
(hybridization between different orbital characters) and bandwidths 
all depend on the size of the A-site cation. 
%%%%

%%%%%%%%%%%%%%%%%%%%
%%% METHODS    
%%%%%%%%%%%%%%%%%%%%
{\it \bf Methods.-} 
The ARPES measurements were conducted at the Synchrotron Radiation Center 
(SRC, University of Wisconsin, Madison) and the \mbox{CASSIOPEE} 
beamline of Synchrotron Soleil (France)
at temperatures \mbox{$T=7~\text{--}~20$~K} and pressures 
lower than \mbox{$P=6\times10^{-11}$~Torr}. 
%%%
Details on the surface preparation and creation of the 2DES
are discussed in the \textcolor{blue}{Supplementary Material}.
%%%
Density Functional Theory (DFT) calculations were carried out on bulk CaTiO$_3$. 
Values of the lattice parameters, tilt angles and band gaps were estimated
and compared to experimental data reported in~\cite{Knight2011}. 
Of the three exchange-correlation functionals tested, the hybrid one (HSE06) 
gave the best agreement with the experimental values 
(see \textcolor{blue}{Supplementary Material} for a comparison of the results 
obtained with different functionals). 
With this choice,
the calculated lattice parameters differ from the experimental estimates 
by less than $0.01$~\AA~and the tilt angles by less than 0.2$^\circ$.
%%%
The band gap is estimated to be 3.62~eV 
as compared to the experimental value of 3.50~eV~\cite{Ueda1999}.
All through this paper, directions and planes are defined in the quasi-cubic cell of CaTiO$_3$. 
In this way, the $(x,y,z)$ axes used to express orbitals and wave functions 
are defined along the Ti-Ti directions.
In contrast, for experimental convenience,
the indices $h$, $k$ and $l$ of $\Gamma_{hkl}$ correspond to
the reciprocal lattice vectors of the orthorhombic unit cell.
%%%

%%%%%%%%%%%%%%%%%
%%% RESULTS
%%%%%%%%%%%%%%%%%
{\it \bf Experimental results.-} 
%%
%%%%%%%%%%%%%%%%%%%%%%%%%%%%%%%%%%%%%
%% Qualitative description of data
%%%%%%%%%%%%%%%%%%%%%%%%%%%%%%%%%%%%%
Figs.~\ref{fig:CTO}(a,~b) show the different observed Fermi surfaces 
in the $(001)$ plane of pseudo-cubic CaTiO$_3$. 
They were measured, respectively, around $\Gamma_{005}$ using $h\nu=57$~eV photons  
with linear vertical (LV) polarization, and around $\Gamma_{115}$ using $h\nu=67$~eV photons
with linear horizontal (LH) polarization. One Fermi sheet consists of a four-pointed star 
as shown in Fig.~\ref{fig:CTO}(a), while two other Fermi sheets are circular 
as seen in Fig.~\ref{fig:CTO}(b).
%%%
Figs.~\ref{fig:CTO}(c,~d) present the energy-momentum maps 
close to the bulk $\Gamma_{005}$ point along the $\langle 010 \rangle$ direction,
using respectively LH and LV polarizations.
In Fig.~\ref{fig:CTO}(d) one observes two dispersive light bands 
and a portion of heavy band close to the Fermi level, 
whereas the other part of the heavy band, with bottom about $62$~meV below $E_F$, 
can be seen in Fig.~\ref{fig:CTO}(c).
%%%

%%%%%%%%%%%%%%%%%%%%%%%%%%%%%%%%%%%%%
%% Hybridization based on comparison with STO
%%%%%%%%%%%%%%%%%%%%%%%%%%%%%%%%%%%%%
To understand the originality of the 2DES in CaTiO$_3$, 
it is instructive to compare its electronic structure with that found in SrTiO$_3$. 
Figs.~\ref{fig:CTO}(e,~f) show, respectively, the Fermi surface and $E-k$ map obtained 
at the Al-capped SrTiO$_3$(001) surface 
--a protocol recently developed by us to create highly homogeneous 2DES 
on several oxides~\cite{Rodel2016}. 
%%%
We thus identify three bands, two light and one heavy, in the $E-k$ maps of both materials. 
In SrTiO$_3$, the two light bands have $d_{xy}$ character, 
while the heavy band has $d_{yz}$ ($d_{xz}$) character
along $k_x$ ($k_y$)~\cite{Santander-Syro2011,Meevasana2011}.
%%%
For CaTiO$_3$, as will be fully justified by our DFT calculations below, 
we also identify the subbands as states of the $t_{2g}$ manifold.
The two light bands correspond to $d_{xy}$ bands forming circular Fermi surface sheets 
(see the \textcolor{blue}{Supplementary Material} for additional data 
close to $\Gamma_{005}$ and $\Gamma_{115}$, 
as well as photon energy dependence of ARPES data in CaTiO$_3$, 
to confirm their orbital and 2D characters). 
However, the dispersion of the heavy band is clearly different in SrTiO$_3$ and CaTiO$_3$. 
The rotation of the oxygen octahedra in CaTiO$_3$ breaks the cubic symmetry 
of the ideal perovskite lattice (SrTiO$_3$) and thus can result 
in the hybridization of orbitals of different azimuthal quantum numbers. 
The hybridization of the $d_{xz}$ and $d_{yz}$ bands in CaTiO$_3$ is evident 
from the star-shaped Fermi surface in Fig.~\ref{fig:CTO}(a),
which can be understood as resulting from the hybridization 
of the two elliptic Fermi surface sheets in Fig.~\ref{fig:CTO}(e),
and is further supported by the non-parabolic dispersion 
as well as the light polarization dependence of the heavy band 
in Figs.~\ref{fig:CTO}(c,~d). 
In fact, the dispersions of the heavy subbands in CaTiO$_3$
can be fitted using a minimal one-layer tight-binding model 
assuming hybridization of the $d_{xz}$ and $d_{yz}$ bands, 
as shown by the red curves in Figs.~\ref{fig:CTO}(a,~c,~d).
% for the specific case of the hybrid heavy band. 
%%
Based on such model, Figs.~\ref{fig:CTO}(g,~h,~i) 
show the momentum-resolved fraction of the orbital character of the hybrid heavy band, 
demonstrating the hybridization between the $d_{xz}$ and $d_{yz}$ orbitals in CaTiO$_3$,
and the pure $d_{xz}$ orbital character of the heavy band in SrTiO$_3$. 
Details on the used tight-binding model are provided 
in the \textcolor{blue}{Supplementary Material}.
%%%

%%%%%%%%%%%%%%%%%%%%%%%%%%%%%%%%%%%%%
%% Quantitative description of data
%%%%%%%%%%%%%%%%%%%%%%%%%%%%%%%%%%%%%
From Fig.~\ref{fig:CTO}(c,~d), the bottoms of the $d_{xy}$ subbands 
at the surface of CaTiO$_3$ are located at $-158$~meV and $-27$~meV, 
while the bottom of the hybrid $(d_{xz}, d_{yz})$ heavy subband is at $-62$~meV. 
Parabolic fits around $\Gamma$ yield an effective mass of approximately $m_{d_{xy}}~=~1.1~m_e$ 
for the $d_{xy}$ bands, and $m_{d_{xz, yz}}~(\Gamma)~=~2.7~m_e$ for the heavy band. 
Based on the tight-binding model described before, 
the mass of the heavy band sufficiently away from $\Gamma$ 
(close to its Fermi momenta $k_F$, where orbital hybridization is negligibly small) 
is $m_{d_{xz, yz}}(k_F)~\approx~15~m_e$. 
The Fermi momenta of the $d_{xy}$ subbands are $0.07$~\AA$^{-1}$ and $0.20$~\AA$^{-1}$, 
and $0.38$~\AA$^{-1}$ for the hybrid heavy subband.
This gives an electron concentration of 
$n_{2D} \approx 1.2\times10^{14} $~cm$^{-2}$, or about $0.17$ electrons per $a^2$, 
where $a$ is the pseudo-cubic lattice constant of the orthorhombic lattice.
%%%

%%%%%%%%%%%%%%%%%
%%% DFT CALCULATIONS
%%%%%%%%%%%%%%%%%
{\it \bf Numerical calculations.-}
We carried out DFT calculations to understand how the rotation of the 
oxygen octahedra surrounding the Ti$^{4+}$ cation, and the concomitant 
altered bonding angle Ti-O-Ti, affect the orbital order of the crystal field split 
$t_{2g}$ $(d_{xy}, d_{yz}, d_{xz})$ and $e_g$ $(d_{z^2}, d_{x^2-ˆ'y^2})$ orbitals 
in bulk CaTiO$_3$.
In fact, as shown by the projected densities of states in Fig.~\ref{figmain1}(a), top panel,
the top of the valence band (set as zero of energy) 
has mostly oxygen character, while Ti states contribute mainly 
to the bottom of the conduction band (CB). 
Moreover, as demonstrated in Fig.~\ref{figmain1}(a), bottom panel, 
a decomposition into Ti~$e_g$ and $t_{2g}$ components shows that the CB minimum 
displays predominantly a $t_{2g}$ character, consistent with the octahedral environment 
of Ti cations. 
Thus, despite the non-negligible tilt of the TiO$_6$ octahedra 
in the bulk CaTiO$_3$ structure, the contribution of the $e_g$ component 
to the bottom of CB is small, and it totally vanishes at the CB minimum at $\Gamma$. 
Thus, the 2DES at the $(001)$ surface of CaTiO$_3$ should be mainly composed 
of the $t_{2g}$ states, which justifies the choice of tight-binding orbitals 
used to fit the experimental data in Fig.~\ref{fig:CTO}.
In fact, as shown in Fig.~\ref{figmain1}(b), the projection of the lower-energy 
conduction state into the experimentally studied pseudo-TiO$_2$ plane 
shows clearly that the electron wave function in this plane presents the symmetry 
of $t_{2g}$ orbitals.
The \textcolor{blue}{Supplementary Material} presents a detailed description 
of our DFT calculations. 
%%%

%%%%%%%%%%%%%%%%%%%%%%%%%%%%%%%%%%%%
%%%% DISCUSSION - COMPARISON ATiO3
%%%%%%%%%%%%%%%%%%%%%%%%%%%%%%%%%%%%
%%%%%%%%%%%%
\begin{figure}[th]
  \begin{center}
   	  \includegraphics[clip, width=0.49\textwidth]{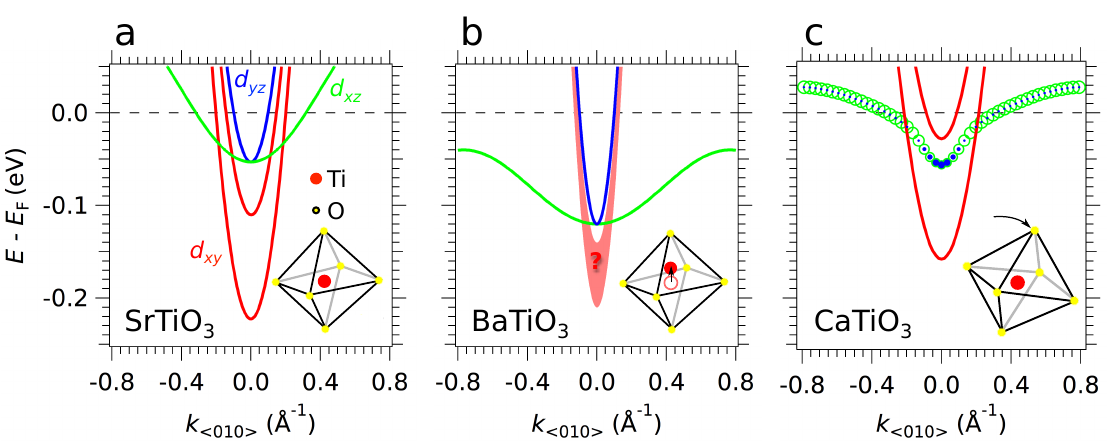}
  \end{center}
  \caption{\label{fig:ATO_comp} 
  		  Oxygen octahedra in ATiO$_3$ (A=Sr,Ba,Ca) perovskites 
  		  and schematic of band dispersion observed in the 2DESs in SrTiO$_3$ (a), 
  		  BaTiO$_3$ (b) and CaTiO$_3$ (c). The black arrows in (b) and (c) 
  		  indicate the distortion occuring in BaTiO$_3$ and CaTiO$_3$. 
  		  The broad red band and the question mark in (b) indicate that the band structure 
  		  of the $d_{xy}$ band was not resolved well by ARPES in BaTiO$_3$. 
  		  The colors of the bands correspond to different orbital characters. 
  		  The size of the blue filled and green empty circles in (c) 
  		  represents the fraction of the $d_{xz}$ (green) and the $d_{yz}$ (blue) 
  		  in the band dispersion.
  		  }
\end{figure}
%%%%%%%%%%%%%
%%
{\it \bf Comparison between various ATiO$_3$ perovskites.-}
The 2DES at the surface of CaTiO$_3$, presented in this paper,
is a new member of the family of ATiO$_3$ perovskites hosting a 2DES on its surface 
(SrTiO$_3$ and BaTiO$_3$)~\cite{Santander-Syro2011,Rodel2016}. 
%%%
The comparison of these 2DES gives insight into the coupling of the electronic 
structure to different lattice symmetries, 
as the three oxides show fundamentally different 
lattice distortions. While SrTiO$_3$ is (close to) the perfect cubic perovskite structure, 
the oxygen octahedra are rotated in CaTiO$_3$, and in BaTiO$_3$ 
the Ti cation moves away from the center of the octahedra resulting in a ferroelectric distortion. 
These rotations/distortions and the corresponding electronic structure of the 2DES's, 
based on our ARPES measurements, are schematized in Figs.~\ref{fig:ATO_comp}(a,~b,~c). 
The ARPES results on the 2DES's are also summarized in Table~\ref{table:ATO_comp}. 
The differences in the electron structure will be discussed in the next paragraphs.
%%%

%%%
\begin{table}[b]
	% \begin{center}
		\centering
		\caption{\label{table:ATO_comp} 
			Relationships between crystal structure and electronic properties
			of the 2DES at the surface of CaTiO$_3$, SrTiO$_3$ and BaTiO$_3$. 
			The Goldschmidt's tolerance factor and the lattice symmetry 
			are given in the first two rows. 
			The effective masses $m^{\star}$ (in units of the free electron mass $m_e$) 
			of the $d_{xy}$ and $d_{xz}$,$d_{yz}$ bands are given in the next two rows. 
			The subsequent four rows give the bottom energies of the $E_{t_{2g}}$ bands 
			together with the energy difference between the two $d_{xy}$ subbands, 
			$\Delta E_{d_{xy}}$. The last row gives the electron density $n_{2D}$ of the 2DES.
			All data correspond to the maximal electron density observed for each of
			the 2DES.
			}
		\begin{threeparttable}
		\begin{tabular}{c | c | c | c }
		\hline \hline
        	&  ~CaTiO$_3$~  & ~SrTiO$_3$~\cite{Rodel2016} & ~BaTiO$_3$~\cite{Rodel2016} \\
    	\hline
			Tolerance & $0.97$ & $1.01$ & $1.08$ \\
			Phase at RT & \small orthorhombic & \small cubic & \small tetragonal \\
			$m^{\star}_{d_{xy}} / m_e$ & $1.1$ & $0.7$ & $0.3 \pm 0.2^{a}$ \\
			$m^{\star}_{d_{xz/yz}} / m_e$ & $2.7$ at $\Gamma$ & $7 \pm 1$ & $10 \pm 2$ \\
			$E_{d_{xy}}^{(1)}$ (meV) & $158$ & $223$ & $200 \pm 60^{a}$ \\
			$E_{d_{xy}}^{(2)}$ (meV) & $27$ & $110$ & -- \\
			$\Delta E_{d_{xy}}$ (meV) & $131$ & $113$ & -- \\
			$E_{d_{xz/yz}}$ (meV) & $62$ & $50$ & $135 \pm 10$ \\
			Orbital order & $xy, (xz/yz), xy$ & $xy, xy, xz/yz $ & $xy, xz/yz$\\  
			$n_{2D}~(10^{14}$cm$^{-2})$ & $1.2$ & $1.4$ & $2.8 \pm 0.4$ \\  
	  	\hline
  		%%%      
		\end{tabular}
	% \end{center}
	%%%
	%\footnotesize{
	\begin{tablenotes}[para]
				  $^{a}$ The dispersion of the light $d_{xy}$ band in BaTiO$_3$ 
				  has not been resolved well by ARPES. 
				  The light electron mass $m^{\star}_{d_{xy}}$ is estimated
				  from the band bottom and Fermi momenta of the $d_{xz}$ or $d_{yz}$
				  band along the ``light'' direction ($x$ for $d_{xz}$, $y$ for $d_{yz}$).
				  The estimated band bottom $E_{d_{xy}}^{(1)}$ is based on the 
				  spectral weight distribution of the $d_{xy}$ band in~\cite{Rodel2016}.
				  %}
	\end{tablenotes}
	\end{threeparttable}
%%%
\end{table}
%%%%

% \textcolor{blue}{Effective masses $m^*$.}
The effective mass of the $d_{xy}$ subbands is larger by a factor of $1.6$ in CaTiO$_3$ 
compared to SrTiO$_3$, due to the rotation-induced decrease 
in the Ti $d$ bandwidth~\cite{Zhong2008d}.
This reduced bandwidth or respectively, increased density of states 
was related to a more robust ferromagnetism at the LaAlO$_3$/CaTiO$_3$ interface 
compared to the LaAlO$_3$/SrTiO$_3$ interface, although the driving force 
for the magnetic order are the  $d_{xz}$ and $d_{yz}$, 
not the $d_{xy}$, orbitals~\cite{Ganguli2014}. 
Due to the orbital hybridization of the $d_{xz}$ and $d_{yz}$ bands, 
$m^*_{d_{xz,yz}}$ is about five times smaller close to $\Gamma$ 
than far away from $\Gamma$ (near $E_F$) in CaTiO$_3$. 
These insights are also of relevance for SrTiO$_3$-based heterostructures, 
as rotations of octahedra can occur at interfaces~\cite{Zhong2008d,Jia2009, Rubano2013}.
%%%
	 
% \textcolor{blue}{Electron densities $n_{2D}$.}
While the electron densities are rather similar in CaTiO$_3$ and SrTiO$_3$ 
(factor of 1.2 ), $n_{2D}$ in BaTiO$_3$ is at least twice as large 
compared to the other oxides. 
The ferroelectric polarization in single domain BaTiO$_3$/SrTiO$_3$ thin films 
is in the upward direction, \emph{i.e.} towards the surface~\cite{Chen2013b}. 
The resulting electric field will influence the confining field of the 2DES and thus, 
the electron density can be altered. Hence, in principle, 
$n_{2D}$ can be controlled by the polarization in the thin film 
which can be manipulated by choosing different substrates~\cite{Chen2013b} 
or by applying strain gradients~\cite{Lu2012e}.
%%%

% \textcolor{blue}{Orbital order.}
The orbital order in 2DESs is mainly determined by the effective mass 
along the confinement direction $m_z^*$~\cite{Santander-Syro2011}. 
As shown previously, the orbital mixing in CaTiO$_3$ influences $m^*_{d_{xz,yz}}$ 
in the surface plane, and will also influence $m_z^*$ for this band. 
Thus, the combined effects of hybridization and electron confinement 
determine the orbital order. Consequently, as seen in Fig.~\ref{fig:ATO_comp} 
and summarized in table~\ref{table:ATO_comp}, the hybridized band in CaTiO$_3$ 
is {\it in between} two $d_{xy}$ subbands, 
in contrast to the $d_{xy}-d_{xy}-d_{xz/yz}$ energy order 
in SrTiO$_3$~\cite{Santander-Syro2011, Rodel2016}.
%%%

The orbital order in the 2DES at the LaAlO$_3$/SrTiO$_3$ interface 
is essential to understand its properties. 
Many of the unusual phenomena at this interface are related to the Lifshitz transition 
occurring at electron densities at which the heavy bands $d_{xz}/d_{yz}$ 
start to be populated~\cite{Joshua2012,Joshua2013, Liang2015}. 
In contrast, other phenomena are only observed in pure $d_{xy}$ systems, 
\emph{e.g.} the Quantum Hall Effect~\cite{Trier2016}. In SrTiO$_3$-based interfaces 
the control of the orbital order and occupancy is based on adjusting the electron density 
and the spatial extension as well as depth of the quantum well 
confining the electrons, depending for example on the composition 
of the oxide heterostructure~\cite{Trier2016}. 
Another way is to chose different surface or interface orientations 
in SrTiO$_3$~\cite{Roedel2014, Herranz2015}. The present study demonstrates 
another possibility by choosing different A-site cations in the perovskite lattice. 
New insights in the properties at the interfaces of complex oxides could be gained 
by studying Lifshitz transition in 2DESs in CaTiO$_3$ and BaTiO$_3$.
%%%

%%%%%%%%%%%%%%%%%
%%%% CONCLUSIONS
%%%%%%%%%%%%%%%%%
{\it \bf Conclusions.-}
We studied the coupling between the lattice structure and the electronic structure 
of 2DES at the surface of three different insulating perovskites ATiO$_3$. 
Our reference system is the 2DES in cubic SrTiO$_3$ which has been intensively studied 
at its bare surface as well as at the LaAlO$_3$/SrTiO$_3$ interface. 
The orthorhombic distortions in CaTiO$_3$ result in a hybridization 
of the $d_{xz}$ and $d_{yz}$ orbitals. The ferroelectric distortions in BaTiO$_3$ 
result in a macroscopic polarization which influences the electron density of the 2DES. 
Moreover, the distortions change band width as well as the orbital order of the $t_{2g}$ 
manifold in the 2DES. Both band width and orbital order 
influence the macroscopic, \emph{e.g.} magnetic and transport,
properties of the 2DESs~\cite{Ganguli2014,Joshua2013}. 
Our results motivate the study of interfaces beyond SrTiO$_3$ 
as so far the question of whether the properties of the LaAlO$_3$/SrTiO$_3$ interface 
can be generalized to 2DES in other perovskite oxides remains largely unanswered.
%%%%%%%%%%%

\acknowledgments 
%%%
% {\it Acknowledgements.-} 
%%%
We thank C\'edric Baumier for help during ARPES experiments,
and Marcelo Rozenberg for discussions.
This work was supported by public grants from the French National Research Agency (ANR), 
project LACUNES No ANR-13-BS04-0006-01, 
the ``Laboratoire d'Excellence Physique Atomes Lumi\`ere Mati\`ere'' 
(LabEx PALM projects ELECTROX and 2DEGS2USE) overseen by the ANR as part of the 
``Investissements d'Avenir'' program (reference: ANR-10-LABX-0039),
and the CNRS-CONICET 2015-2016 collaborative project AMODOX (project number:~254274).
T.~C.~R. thanks funding from the RTRA--Triangle de la Physique (project PEGASOS).
R.~W. acknowledges support from CONICET (grant No. PIP 114-201101-00376), 
and ANPCyT (grant No. PICT-2012-0609).
M.~G. and A.F.S.-S. acknowledge the support received from the Institut Universitaire de France.
%%%%%%%%%%%%%%%%%%%%%%%%%%%%%%%%%%%%%%%%%%%%%%%%%%%%%%%%%%%%%%%%%%%%%%%%%%%%%%%%%%%%%%%
%%%%%%%%%%%%%%%%%%%%%%%%%%%%%%%%%%%%%%%%%%%%%%%%%%%%%%%%%%%%%%%%%%%%%%%%%%%%%%%%%%%%%%%

%%%%%%%%%%%%%%%%%%%%%%%%%%%%
%%%% SUPPLEMENTARY MATERIAL
%%%%%%%%%%%%%%%%%%%%%%%%%%%%
\section*{SUPPLEMENTARY MATERIAL}

% \setcounter{figure}{0}
% \setcounter{table}{0}
% \makeatletter 
% \renewcommand{\thefigure}{S-\arabic{figure}}
% \renewcommand{\thetable}{S-\arabic{table}}
% \renewcommand\@biblabel[1]{[SR#1]}
%table of contents
%\begin{itemize}
%	\item surface preparation
%	\item creation of oxygen vacancies
%	\item DFT calculations
%	\item photon energy dependence
%	\item spatial extension of 2DES
%	\item complete data set $\Gamma_{005}$ and $\Gamma_{115}$
%	\item tight-binding model
%\end{itemize}

\subsection*{Surface preparation}
%%%
\begin{figure}[b]
  \begin{center}
   	  \includegraphics[clip, width=0.3\textwidth]{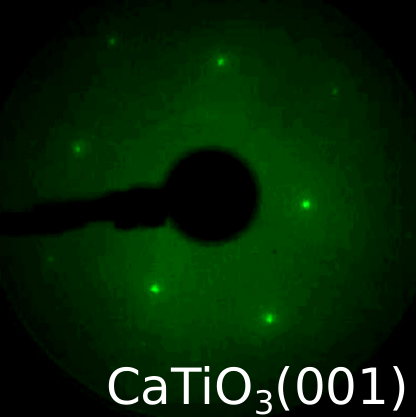}
  \end{center}
  \caption{\label{fig:CTO_LEED} 
  		   LEED image of a polished CaTiO$_3$(001) surface after annealing 
  		   at approximately $T=550\text{--}600^{\circ}$C in UHV. 
  		   The kinetic energy of the incident electrons was set to 81~eV.
  		   }
\end{figure}
%%%
Single crystals of CaTiO$_3$ of $5\times5\times0.5$~mm$^3$ were provided by SurfaceNet GmbH. 
To clean the polished surface, the crystal was annealed in ultra high vacuum 
at a temperature of about 550-600$^\circ$C, 
as at this temperature most carbon-based compound desorb from the surface 
of the closely related perovskite SrTiO$_3$~\cite{Kawasaki1996}. 
The resulting low energy electron diffraction (LEED) image in Fig.~\ref{fig:CTO_LEED} 
shows an unreconstructed surface. 
Alternatively, single crystals were fractured at low temperatures $T=7\text{--}20$~K 
in ultra high vacuum. The fractured surfaces were not characterized by electron diffraction. 
However, the periodicity of the electronic structure from ARPES measurements 
demonstrates the surface crystallinity and the absence of surface reconstructions. 
%%%%%   

\subsection*{Creation of oxygen vacancies}
\begin{figure}[t]
  \begin{center}
   	  \includegraphics[clip, width=0.48\textwidth]{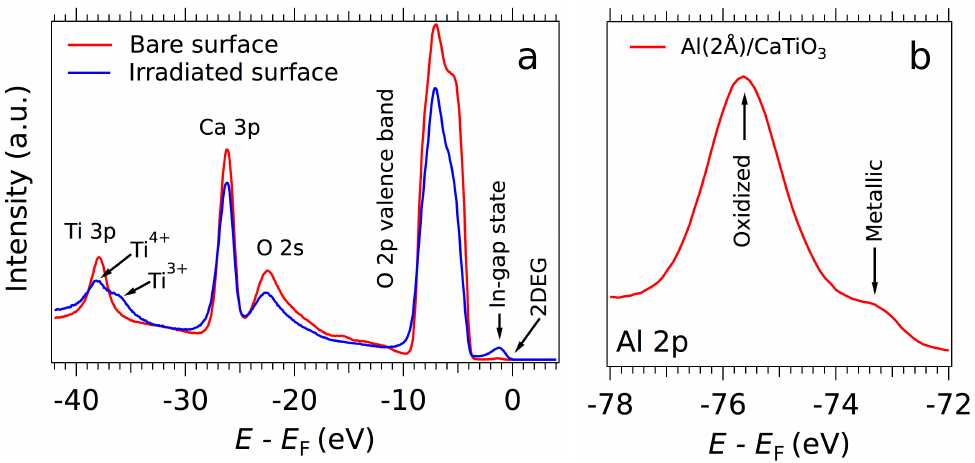}
  \end{center}
  \caption{\label{fig:CTO_ox} 
  		  (a) Angle-integrated spectra of a fractured CaTiO$_3$(001) surface 
  		  measured at a photon energy of \mbox{$h\nu=100$~eV} showing the partial density of states 
  		  for binding energies between -42~eV and 2~eV. 
  		  The red curve was measured right after the first exposure to light 
  		  whereas the blue one was measured about an hour later. 
  		  The creation of oxygen vacancies is evidenced by the Ti$^{3+}$ shoulder 
  		  of the Ti~$2p$ peak and the presence of the in-gap state as well as the 2DES. 
  		  (b) Aluminium $2p$ peak measured at $h\nu=100$~eV after depositing 2~\AA~of aluminium 
  		  on a clean and polished CaTiO$_3$(001) surface. 
  		  The part of the peak corresponding to oxidized Al demonstrates the occurrence 
  		  of a redox reaction at the interface of Al and oxide.
  		  }
\end{figure}
%%%

%% \textcolor{blue}{\textit{Changes in DOS and localized vs itinerant.}}
As in previous works, the 2DES is formed by the creation of oxygen vacancies 
due to desorption induced by UV irradiation~\cite{Walker2015} 
or due a redox reaction with Al~\cite{Rodel2016}. 
Fig.~\ref{fig:CTO_ox}(a) shows a comparison between the stoichiometric surface (red curve) 
and the surface reduced by the synchrotron beam (blue curve)
after the oxygen depletion reached a saturation value 
at the given temperature and photon flux. 
Oxygen vacancies produce an electron transfer from oxygen to Ti, 
as evidenced by the shoulder of the Ti~$3p$ peak corresponding to Ti$^{3+}$ cations,
and the in-gap state as well as the 2DES formed close to the Fermi level $E_F$. 
As in other reduced oxide surfaces, the in-gap state corresponds to electrons 
localized at Ti cations close to the vacancy whereas the 2DES represents 
itinerant electrons~\cite{Jeschke2014, Roedel2015}. 
This dual character of excess electrons and the interplay between the two 
is essential to understand the magnetic properties of SrTiO$_3$ based 
interfaces and surfaces~\cite{Lechermann2016}.

%% \textcolor{blue}{\textit{Vacancy concentration}}
To roughly estimate the concentration of oxygen vacancies at the surface, 
the Ti~$3p$ peak of the blue curve in Fig.~\ref{fig:CTO_ox}(a) 
is fitted by two Voigt peaks, corresponding to the Ti$^{4+}$ and Ti$^{3+}$ contributions, 
and a Shirley background. The fraction of the area of the Ti$^{3+}$ Voigt peak 
is about 36\% compared to the total area of the Ti~$3p$ peak. 
Assuming that an oxygen vacancy results in an transfer of two electrons to Ti, 
the vacancy concentration is about $x=6\%$ in CaTiO$_{3(1-x)}$. 
Note that this value corresponds to the weighted average over the probing depth 
defined by inelastic mean free path of the photoelectrons in the solid. 
The concentration of oxygen vacancies varies depending on photon flux, 
exposure time and temperature. 
The given value was obtained at a surface in which the effect of oxygen desorption 
saturated under the given experimental conditions.

%% \textcolor{blue}{\textit{Bend bending}}
The position of the Ti$^{4+}$ peak fitted by a Voigt shape shifts by 230~meV 
in binding energy between the blue and red curve in Fig.~\ref{fig:CTO_ox}(a). 
This value corresponds to the minimal band bending at the oxygen-deficient surface of CaTiO$_3$ 
shown in Fig.~\ref{fig:CTO_ox}(a), as it is again a weighted average over the probing depth 
and a small band bending might already be present at the first measurement (red curve).

%% \textcolor{blue}{\textit{Oxidized Al 2p}}     
Alternatively to the synchrotron irradiation, oxygen vacancies are created 
by the deposition of 2~\AA~of aluminum on the clean and polished surface of CaTiO$_3$ 
at temperatures of about $T=50-100^\circ$C. 
Details on the Al deposition are described elsewhere~\cite{Rodel2016}. 
As shown in Fig.~\ref{fig:CTO_ox}(b), a large part of the deposited aluminum 
is oxidized as evidenced by the peak at about $-75.6$~eV, 
and only a small part is still metallic (peak at about $-73$~eV).
%%%

%%%%%%%%%
\subsection*{DFT: Computational method and settings}
%%%
Most of the calculations were performed within the Density Functional Theory (DFT), 
implemented in VASP (Vienna ab initio simulation package)~\cite{Kresse1996, Kresse1993}. 
Gradient-corrected GGA (PW91)~\cite{pw91}, GGA+U~\cite{DudarevPRB1998} (U$=5.0$~eV, J$=0.64$~eV), 
and hybrid (HSE06)~\cite{hse06a,hse06b} exchange-correlation functionals were used 
to give robust grounds for the description of the electronic structure of the bulk. 
The interaction of valence electrons with ionic cores was described 
within the projector augmented wave (PAW) method~\cite{Blochl1994, Kresse1999}, 
and the Kohn-Sham orbitals were developed on a plane-wave basis set. 
Standard Calcium and Titanium (including 3$p$ electrons in the valence band), 
and soft oxygen (energy cutoff of 270~eV) pseudopotentials 
provided with VASP were used~\cite{Blochl1994, Kresse1999}, 
enabling a full structural relaxation of all considered systems at the hybrid level. 
The applicability of soft oxygen pseudopotential was validated on the GGA+U level, 
with results obtained with both the soft and full (energy cutoff of 400~eV) 
oxygen pseudopotentials. Table~\ref{tab1} shows a perfect agreement 
of the calculated structural (lattice parameters and tilt angles) and electronic (band gap) 
characteristics of bulk CaTiO$_3$ obtained with the two types of pseudopotentials.
%%%%

%%%
In CaTiO$_3$, the Ca$^{2+}$ cations are too small to fit in a undistorted 
cubic perovskite which results in a $a^-a^-c^+$ rotation 
(Glazer notation~\cite{Glazer1972}) 
of the oxygen octahedra surrounding the Ti$^{4+}$ cation and hence, 
an altered bonding angle Ti-O-Ti. 
For the bulk calculations we used such orthorhombic unit cell
% the orthorhombic $Pnma$ CaTiO$_3$ unit cell 
% with the $a^-a^-c^+$ tilt pattern and 
optimizing its shape and volume 
until all elements of the stress tensor were smaller than 0.01~eV/$\text{\AA}^3$. 
Simultaneously, positions of all atoms were optimized so as to make all forces 
less than 0.01~eV/$\text{\AA}$. With these settings a 
$(6 \times 6 \times 4)$ $\Gamma$-centered Monkhorst Pack sampling 
of the bulk Brillouin zone assures a convergence of calculated energy differences 
to $0.02$~eV/CaTiO$_3$, and of lattice parameters to less than 0.01~\AA. 
%%%
We have done some calculations using the Wien2k package~\cite{Wien2k}, 
in order to double check our results.
In particular, charge density plot showed in 
% Fig.~\ref{figmain1}(c)
Fig.~2(c) of the main text were done using this code.

%%%%%
\subsection*{DFT: Bulk results for different exchange-correlation functionals}
Table~\ref{tab1} summarizes the results obtained for bulk CaTiO$_3$ 
at different levels of approximation. 
Pure GGA predicts a bulk structure (lattice parameters and tilt angles) 
in relatively good agreement with the experimental data, 
but largely underestimates the band gap. 
While the GGA+U approximation considerably improves the band gap, 
it does not match closely the experimental result despite the large $U-J$ value 
used in the calculations.
Moreover, GGA+U introduces an non-negligible overestimation 
of the lattice parameters and tilt angles. 
By contrast, the hybrid approach produces an excellent agreement 
between the calculated and experimental results in what concerns 
both the atomic structure and the electronic characteristics.
%%%

%%%%%%
% Table comparison DFT functionals
%%%%%%
\begin{table}
\begin{ruledtabular}
\begin{tabular}{ l|c c c c c }
& Exp. & GGA & GGA+U  & GGA+U(s) & HSE(s)\\
\hline
a ($\text{\AA}$)&  5.36 & 5.40 &5.43 &  5.43 & 5.36 \\
b ($\text{\AA}$)&  5.43 & 5.50 &5.55 &  5.55 & 5.44 \\
c ($\text{\AA}$)&  7.62 & 7.68 &7.74 &  7.73 & 7.62 \\
$\phi$ ($^\circ$)&   12 & 12.1 & 13.7 & 13.9 & 12.3 \\
$\theta$ ($^\circ$) & 9 &  9.2 &  9.7 &  9.8 &  8.8 \\
Gap (eV)        & 3.50  & 2.28 & 2.90 & 2.90 & 3.62 \\
\end{tabular}
\end{ruledtabular}
\caption{Experimental and calculated structural and electronic characteristics 
		 of bulk CaTiO$_3$: lattice parameters $a, b, c$ (\AA), 
		 tilt angles $\theta, \phi$ ($^\circ$), and band gap (eV), 
		 obtained from GGA, GGA+U and HSE calculations. 
		 (s) denotes results obtained with the soft oxygen pseudopotential.
		 }
\label{tab1}
\end{table}
%%%%%%%%%%%%
%%%%%%
% Figure comparison DGT functionals
%%%%%%
\begin{figure*}
   	\begin{center}
   	  \includegraphics[clip, width=0.95\textwidth]{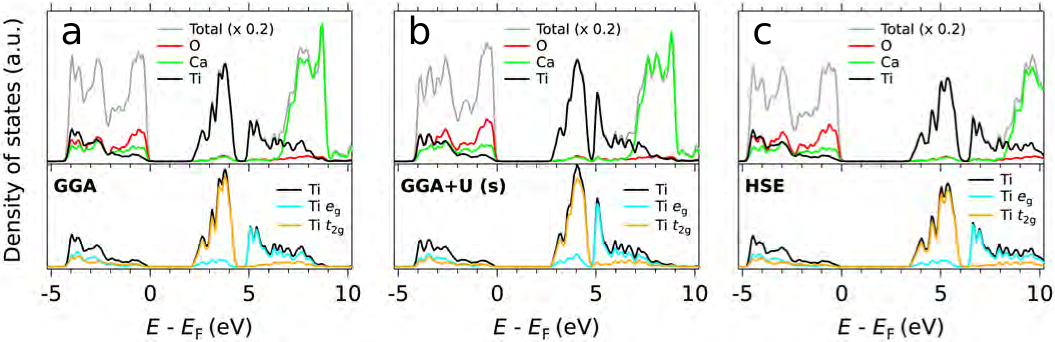}
  	\end{center}
	\caption{Total, Ti-, Ca- and O-projected densities of states of bulk CaTiO$_3$ 
		 obtained from standard GGA (a), 
		 from GGA+U with soft oxygen pseudopotential (b), 
		 and from HSE calculations with soft oxygen pseudopotential (c).
		 The bottom curves of each panel show the decompositions 
		 of the Ti-projected density of states
		 into Ti $e_g$ ($d_{z^2} + d_{x^2 - y^2}$) 
		 and Ti $t_{2g}$ ($d_{xy} + d_{yz} + d_{xz}$) components.
		 }
\label{fig1}
\end{figure*}
%%%%%%%%%%%%%

%%%
Projected densities of states are reported in Fig.~\ref{fig1}. 
Regardless of the level of approximation, GGA, GGA+U or HSE 
and the type of the oxygen pseudoptential used, the top of the valence band 
has mostly an oxygen character, while Ti states contribute mainly 
to the bottom of the conduction band. 
Decomposition into Ti $e_g$ ($d_{z^2} + d_{x^2 - y^2}$) 
and $t_{2g}$ ($d_{xy} + d_{yz} + d_{xz}$) components,
obtained by a rotation of the unit cell to the axes of the conventional cubic unit cell,
%%
% obtained with $x$ and $y$ axes of the coordinates system within the CaTiO$_3$(001) plane 
% and the $x$ axes along the diagonal of the surface unit cell of the orthorhombic lattice,
%% 
shows that conduction band minimum displays mainly a $t_{2g}$ character, 
consistent with the octahedral environment of Ti cations. 
Despite the non-negligible tilt of the TiO$_6$ octahedra in the bulk CaTiO$_3$ structure, 
the contribution of the $e_g$ component to the bottom of conduction band is small, 
and it totally vanishes at the CB minimum at $\Gamma$. 
We note that hybrid calculations systematically predict larger bandwidths, 
as illustrated by the width of the $t_{2g}$ component in the bottom part of conduction band, 
equal to more than 2.5~eV in HSE and to about 2~eV in GGA+U. 
The effect of band narrowing is likely linked to the overestimation 
of the lattice parameters and inter-atomic distances by the GGA+U approximation. 
We also note that the GGA+U approximation overestimates the tilt of the angles 
and predicts a larger tilt than either GGA or hybrid approaches, and would thus lead 
to a larger hybridization of the $t_{2g}$ and $e_{g}$ states. 
This feature can also be seen in Fig.~\ref{fig1} 
where the percentage of the $e_g$ is higher at the bottom of the conduction band 
as compared to GGA or hybrid calculations.
%%%%%

%%%%%
\subsection*{Photon energy dependence}
%%%
\begin{figure*}%[b]
  \begin{center}
   	  \includegraphics[clip, width=0.9\textwidth]{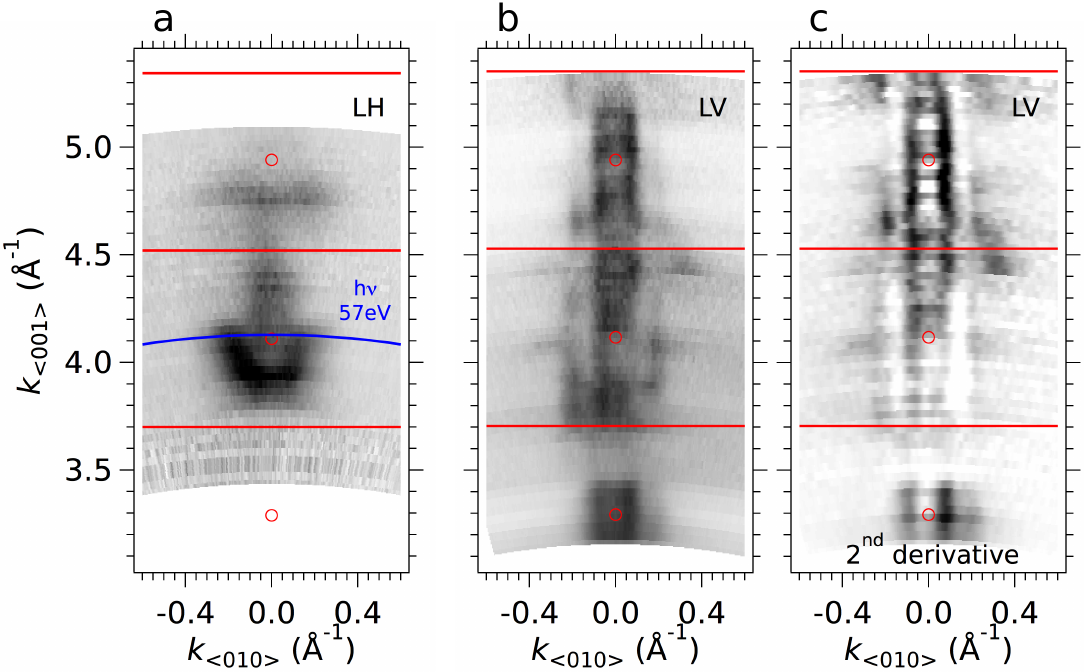}
  \end{center}
  \caption{\label{fig:CTO_hv}  
			2DES at the (001) surface of CaTiO$_3$.
			(a) Fermi surface map in the $k_{\langle001\rangle}-k_{\langle010\rangle}$ plane 
				measured by changing the photon energy between $h\nu=37$~eV and $h\nu=90$~eV 
				using linear horizontal polarization. 
				The $\langle001\rangle$ direction is the confinement direction 
				perpendicular to the surface. 
				To relate the photon energy $hv$ to momentum $k_{\langle001\rangle}$ 
				the inner potential was set to $V_0=12$~eV. 
				The red lines show the Brillouin zone border of the orthorhombic lattice 
				and the red markers correspond to $\Gamma$ points of the reciprocal lattice. 
				The blue curve corresponds to the cut in reciprocal space at $h\nu=57$~eV 
				as shown in the $E_k$ maps in 
				% Fig.~\ref{fig:CTO}(c,d) 
				Figs.~1(c,~d) of the main text.
			(b) Fermi surface map with photon energies ranging between $h\nu=30$~eV 
				and $h\nu=100$~eV using linear vertical polarization. 
			(c) Same as (b) but based on $2^{nd}$ derivatives of the $E-k$ maps. 
				All shown maps are averaged intensities over an energy range 
				of at most $E_F \pm 10$~meV. 
  		}
\end{figure*}
%%%

%%%
The measurement of the photon energy dependence is a way to confirm 
the confined character of the electronic states, 
as changing the photon energy corresponds to probing the electronic structure 
along the confinement direction or respectively, perpendicular to the surface. 
Based on Heisenberg's principle, confinement in real space 
results in a large uncertainty of the momentum, \textit{i.e.} non-dispersing bands 
along the confinement direction in reciprocal space. 
Previous studies on the 2DES in perovskites demonstrated that $d_{xy}$ orbitals 
form tubular Fermi surfaces along the confinement direction 
due to their 2D character~\cite{Plumb2014,Rodel2016}. 
In contrast, the bands of $d_{xz}$ and $d_{yz}$ bands disperse, 
forming neither purely 2D nor strictly 3D states~\cite{Plumb2014,Rodel2016}.
%%%

%%%
Fig.~\ref{fig:CTO_hv} shows that similar dispersions can be observed 
in the 2DES at the (001) surface of CaTiO$_3$. Using linear horizontal (LH) polarization 
the band of mixed $(d_{xz},d_{yz})$ orbital character can be probed close to normal emission 
(compare to 
% Fig.~\ref{fig:CTO}(a,~c)
Figs.~1(a,~c) 
in the main text measured at $h\nu=57$~eV). 
As displayed by the Fermi surface map in Fig.~\ref{fig:CTO_hv}(a), 
this band disperses along the confinement direction and is thus not strictly 2D. 
Using linear vertical polarization the light bands shown in 
% Fig.~\ref{fig:CTO}(d)
Fig.~1(d) of the main text can be measured. As evident from Fig.~\ref{fig:CTO_hv}(b,c) 
these bands are non-dispersing perpendicular to the surface. 
\emph{Their non-dispersing character confirms the $d_{xy}$ character of these two bands}. 
Note that the ARPES intensity measured at different photon energies 
was normalized in Fig.~\ref{fig:CTO_hv} to enhance the visibility 
of the dispersing bands at different $k_{\braket{001}}$ values.

%%%%%
% \subsection{Spatial extension of 2DES at the surface of CaTiO$_3$}
%%%%%
% As in previous studies~\cite{Santander-Syro2011}, the spatial extension of the 2DES 
% can be estimated assuming that a triangular-shaped potential well confines the electrons. 
% \textcolor{red}{TO BE COMPLETED:
% The energy splitting of the two $d_{xy}$ orbitals of $\Delta E_{d_{xy}}=131$~meV 
% corresponds to a potential well of length $d=???$ and a confining field of $F=???$. 
% The effective mass of $d_{xy}$ orbitals along the confinement direction is $m_z=???$ 
% based on the DFT calculations on bulk CaTiO$_3$.
% }
%%  

%%%%%
\subsection*{Complete data set measured close to $\Gamma_{005}$ and $\Gamma_{115}$}
%%%%%
\begin{figure*}%[b]
  \begin{center}
   	  \includegraphics[clip, width=0.9\textwidth]{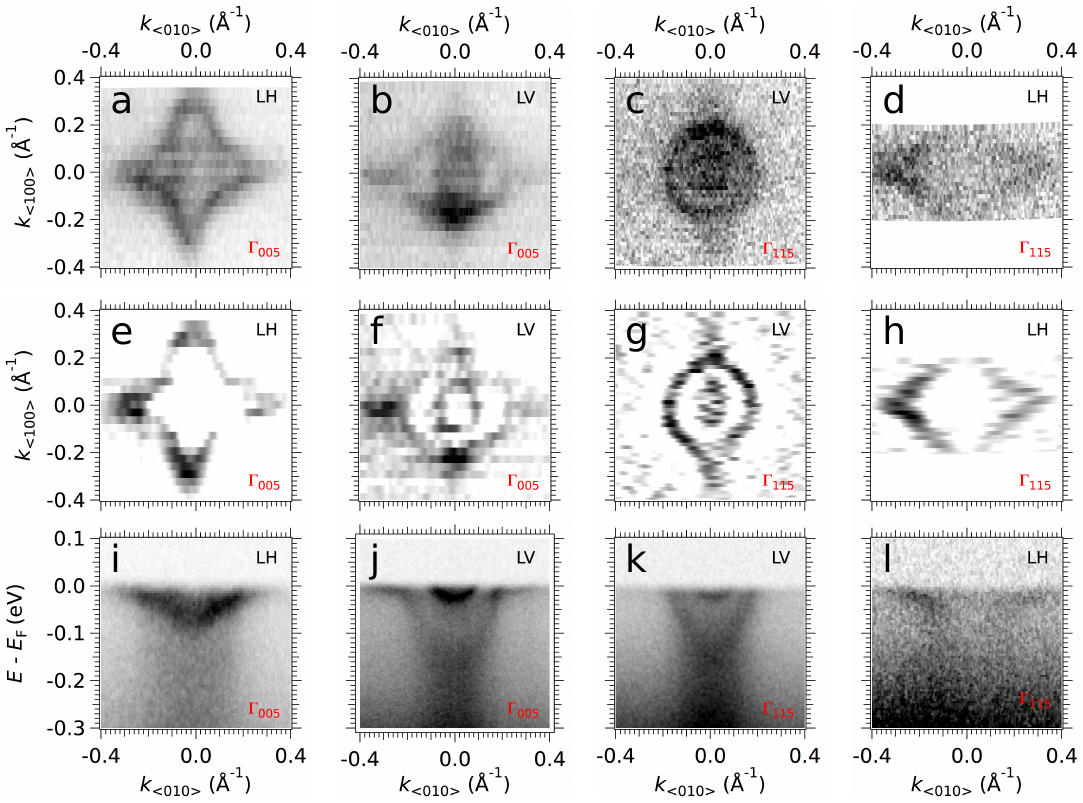}
  \end{center}
  \caption{\label{fig:CTO_comp}  
  		(a-d) Fermi surface intensity maps measured at the surface of CaTiO$_3$(001) 
  			  close to $\Gamma_{005}~(h\nu=57$~eV) (a,b) 
  			  and close to $\Gamma_{115}~(h\nu=67$~eV) (c,d) 
  			  using LH and LV polarization. 
  			  The measurements in (a,b) were conducted on a fractured surface of CaTiO$_3$, 
  			  whereas in (c,d) the Al(2\AA)/CaTiO$_3$ interface was probed. 
  			  The maps shown are averaged intensities over an energy range 
  			  of at most $E_f \pm 10$~meV. 
			(e-h) Same as (a-d) but based on the $2^{nd}$ derivatives of the $E-k$ maps.
			(i-l) Corresponding $E-k$ intensity maps through the $\Gamma$ point.
  		}
\end{figure*}
%%%

%%%
The ARPES data in Fig.~\ref{fig:CTO_comp} show the Fermi surface 
and $E-k$ maps measured close to $\Gamma_{005}$ and $\Gamma_{115}$, 
as well as their $2^{nd}$ derivatives. 
The data close to $\Gamma_{005}$ was measured at $hv=57$~eV at a fractured (001) surface, 
the one close to $\Gamma_{115}$ at $hv=67$~eV at the interface 
between oxidized Al and CaTiO$_3$(001). 
Figs.~\ref{fig:CTO_comp}(a,~c,~i,~j) are identical to the 
% Figs.~\ref{fig:CTO}(a,~b,~c,~d) 
Figs.~1(a,~b,~c,~d)
of the main text. The more complete data set in Fig.~\ref{fig:CTO_comp} 
confirms the conclusions in the main text: the presence of a star-shaped
and two circular Fermi surface sheets or respectively, 
the presence of a heavy band of hybrid $(d_{yz}, d_{xz})$ character
and two light $d_{xy}$ bands. 
The shape of the smallest Fermi sheet formed by the upper light band
is not unambiguously circular from the Fermi surfaces presented 
in Figs.~\ref{fig:CTO_comp}(b,~c,~f,~g). 
However, from Figs.~\ref{fig:CTO_comp}(i,~j), it is evident that 
the polarization-dependent selection rules of the upper light band
are similar to those of the lower light band, 
\emph{i.e.} silenced for LH polarization and enhanced for LV polarization,
as indeed expected for a band of essentially pure $d_{xy}$ character 
in the experimental geometry of the data presented here~\cite{Santander-Syro2011,Rodel2016}.
Together with the non-dispersing character of this band along the confinement 
direction $\braket{001}$, as shown in Fig.~\ref{fig:CTO_hv}(b,c), 
this indicates a $d_{xy}$ character for the upper light band.
As discussed in 
% Fig.~\ref{figmain1}(a) 
Fig.~2(a)
of the main text, 
there might be small contributions of $e_g$ states to the bands forming the 2DES.
%%%

%%%
Note that the electron density of the 2DES at the Al/CaTiO$_3$ interface 
is slightly lower than at the bare surface as evidenced by the smaller Fermi momenta 
and binding energies of the band bottom of the $d_{xy}$ bands 
in Figs.~\ref{fig:CTO_comp}(g,k) compared to Figs.~\ref{fig:CTO_comp}(f,j). 
The given values of Fermi momenta, binding energies and electron densities 
in the main text correspond to values at the bare surface. 
The most probable reason for the difference are the different techniques 
used to prepared the surface and/or create the 2DES.   
%%% 

%%%%%%%%%%%%%%%%%%%%%%%%%%%%%%%%%
\subsection*{Tight-binding model}
%%%%%%%%%%%%%%%%%%%%%%%%%%%%%%%%%
%%%%%%
\begin{figure*}%[b]
  \begin{center}
   	  \includegraphics[clip, width=0.9\textwidth]{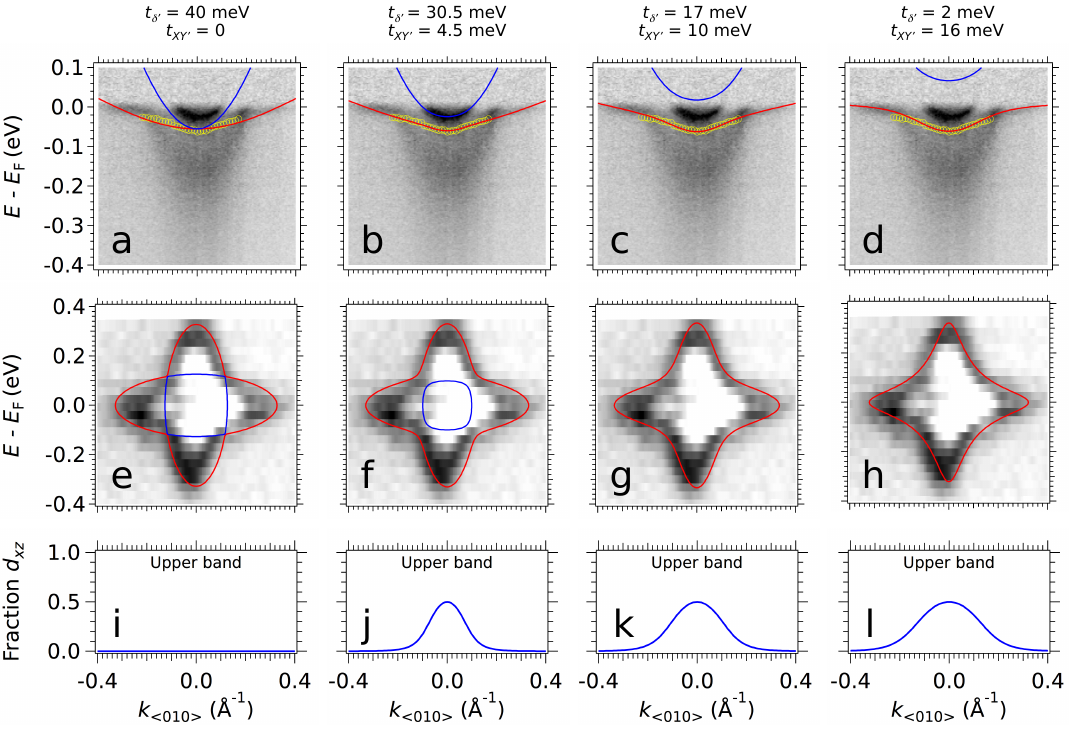}
  \end{center}
  \caption{\label{fig:TB_comp}  
  		(a-d) ARPES $E-k$ intensity maps of the 2DES at the surface CaTiO$_3$(001) 
  		close to $\Gamma_{005}$, measured with LV photons at $h\nu=57$~eV. 
  		The yellow markers are the experimental peak positions of the hybridized heavy band 
  		measured using LH photons at the same energy 
  		--see 
  		% Fig.~\ref{fig:CTO}(c)
  		Fig.~1(c) of the main text.
  		The red and blue curves are tight-binding models of the $d_{xz}/d_{yz}$ doublet
  		for different hopping amplitudes between identical orbitals 	along $\delta$ bonds
  		($t_{\delta'}$) and different hybridization energies 
  		between $d_{xz}$ and $d_{yz}$ orbitals (	$t_{XY'}$), 
  		as specified on top of each panel. 
  		In all cases, the parameters are set to provide the best fit 
  		to the heavy hybridized band for a given $t_{XY'}$.
  		(e-h) Fermi surface of the 2DES at the surface CaTiO$_3$(001) 
  		close to $\Gamma_{005}$, measured with LH photons at $h\nu=57$~eV,
  		to enhance the Fermi surface formed by the hybridized heavy band
  		--as discussed in 
  		% Fig.~\ref{fig:CTO} 
  		Fig.~1 of the main text.
  		The red and blue curves are the tight-binding Fermi surfaces
  		from the corresponding models in panels (a-d).
  		(i-l) Fraction of $d_{xz}$ orbital character along  $k_{<010>}$ (\emph{i.e.} $k_y$) 
  		for the upper band in the hybridized $d_{xz}/d_{yz}$ doublet. 
  		}
\end{figure*}
%%%%%%

We use a simple tight-binding model to rationalize the orbital hybridization 
between $t_{2g}$ orbitals in the 2DES of CaTiO$_3$.
We consider the specific case of the $d_{yz}$ and $d_{xz}$ orbitals giving rise
to the hybrid heavy band seen in 
% Figs.~\ref{fig:CTO}(a,~c) 
Figs.~1(a,c) of the main text.
To reproduce the experimental dispersions and Fermi surfaces
it is sufficient to consider the electron hopping between $d_{yz}$ and $d_{xz}$ orbitals 
on a 2D square lattice. 
The Hamiltonian $H_{XY}$ of the system in the basis $\{d_{I}\}$, 
where $I=(X,Y)$ corresponds to the orbital character $(yz,zx)$ of the two $d$ orbitals, 
is given by:
%%%
\begin{equation*}
H_{XY}=
		\begin{pmatrix}
        	d^\dagger_{X} \\
        	d^\dagger_{Y} \\
		\end{pmatrix}
			^T
		\begin{pmatrix}
			\epsilon_{X} & \epsilon_{XY} \\
			\epsilon_{XY}^\dagger & \epsilon_{Y} \\
		\end{pmatrix}
		\begin{pmatrix}
        	d_{X} \\
        	d_{Y} \\
      	\end{pmatrix},
\label{eq:TB_bulk}
\end{equation*}
%%%
with
%%%
\begin{align*}
    \epsilon_{X}=&-2t_{\pi'}\cos(a~k_y)-2t_{\delta'}\cos(a~k_x)\\
    \epsilon_{Y}=&-2t_{\pi'}\cos(a~k_x)-2t_{\delta'}\cos(a~k_y)\\
		\epsilon_{XY}=&-2t_{XY'}(\cos(a~k_x)+\cos(a~k_y)).\\
	%\label{eq:bulk_eps}
\end{align*}
%%%

%%%
Here, $\epsilon_{X}$ and $\epsilon_{Y}$ describe the intra-layer hopping 
between identical orbitals, whereas $\epsilon_{XY}$ correspond to hopping 
between different orbital characters. The hopping between nearest neighbors of Ti cations 
is characterized by the hopping amplitudes $t_{\pi'}$ and $t_{\delta'}$, 
the hopping between the $d_{xz}$ and $d_{yz}$ orbitals by $t_{XY'}$. 
The Greek letters in the indices of the hopping amplitudes 
correspond to the type of chemical bonding ($\pi,\delta$). 
The tight-binding fit for the hybrid heavy band in CaTiO$_3$, shown in 
% Figs.~\ref{fig:CTO}(a,~c,~d) 
Figs.~1(a,~c,~d) of the main text, 
is based on $t_{\pi'}=0.24$~eV, $t_{\delta'}=0.017$~eV, and $t_{XY'}=0.010$~eV. 
The hopping amplitude $t_{\pi'}$ is defined by the effective mass $m^{\star}_{d_{xy}}$
of the $d_{xy}$ band near $\Gamma$: 
$t_{\pi}=\hbar^2/(2m^{\star}_{d_{xy}} a^2)$, where 
% $m_e$ is the electron mass and 
$a \approx 3.82$~\AA~ the lattice constant of the quasi-square lattice. 
$t_{\delta'}$ and $t_{XY'}$ are adjusted to fit simultaneously the experimental 
band dispersion and Fermi surface of the hybrid heavy band. 
In the case of SrTiO$_3$, $t_{\pi'}$ is calculated 
based on $m^{\star}_{d_{xy}}=0.7 m_e$ in 
% table~\ref{table:ATO_comp} 
table~I of the main text ($m_e$ is the free electron mass), 
and $t_{\delta'}$ based on $m^{\star}_{d_{xz/yz}}\approx 7 m_e$. 
%%%%%%%%%%%

%%%%%%%%%%%%%%%%%%%%%%%%%%%%%%%%%%%
%%% About the TB model ambiguities
%%%%%%%%%%%%%%%%%%%%%%%%%%%%%%%%%%%
Note that the tight-binding parameters to fit the electronic structure of the 2DES
in CaTiO$_3$ are not unambiguous --the main reason being the simplicity of the model. 
Fig.~\ref{fig:TB_comp} shows fits to the hybrid heavy band and Fermi surface
for different sets of parameters, from the case of no hybridization 
between the $d_{xz}$ and $d_{yz}$ orbitals, Figs.~\ref{fig:TB_comp}(a,~e,~i), to the case
of a hybridization energy of $16$~meV, Figs.~\ref{fig:TB_comp}(d,~h,~l).
Comparing Figs.~\ref{fig:TB_comp}(a,~e) with (b,~f), (c,~g) or (d,~h),
it is clear that a certain hybridization between the $d_{yz}$ and $d_{xz}$ orbitals 
is necessary to reproduce the dispersion and Fermi surface of the hybrid heavy band
--the fits becoming better as the hybridization energy approaches $16$~meV.
% the degree of the hybridization cannot be uniquely determined from the quality of the fits.
%%
However, considering only the tight-binding model 
does not clarify the origin of the upper light band forming the inner quasi-circular
Fermi sheet. 
In one case, as shown in Figs.~\ref{fig:TB_comp}(b,~f),
such light band would correspond to the upper band of the pair formed 
by the hybridized $d_{xz}$ and $d_{yz}$ bands. 
In the other case, as shown in Figs.~\ref{fig:TB_comp}(c,~d,~g,~h), 
the upper band of the hybridized doublet would be unoccupied, 
as the splitting is larger due to a larger hybridization energy,
and the experimental upper light band would be the second quantum-well state
of the $d_{xy}$ band.
As stated in the main text, and discussed previously in this Annex,
we assigned $d_{xy}$ character to the upper light band 
due to both its non-dispersing character along the surface normal, 
as shown in Figs.~\ref{fig:CTO_hv}(b,c), and its polarization-dependent selection rules,
Figs.~\ref{fig:CTO_comp}(i,~j), 
which are essentially the same as those of the lower light band.  
%%

%%%%%%%%%%%%%%%%%%%%%%%%%%%%%%%%%%%%%%%%%%%%%%%%%%%%%%%%%%%%%%%%%%%%%%%%%%%%%%%%%%%%%%%
%%%%%%%%% REFERENCES MS+SupMat

% \bibliographystyle{unsrt}
% \bibliographystyle{apsrev4-1}
%%\bibliography{MyCollection.bib}
% \bibliography{}

\begin{thebibliography}{36}%
\makeatletter
\providecommand \@ifxundefined [1]{%
 \@ifx{#1\undefined}
}%
\providecommand \@ifnum [1]{%
 \ifnum #1\expandafter \@firstoftwo
 \else \expandafter \@secondoftwo
 \fi
}%
\providecommand \@ifx [1]{%
 \ifx #1\expandafter \@firstoftwo
 \else \expandafter \@secondoftwo
 \fi
}%
\providecommand \natexlab [1]{#1}%
\providecommand \enquote  [1]{``#1''}%
\providecommand \bibnamefont  [1]{#1}%
\providecommand \bibfnamefont [1]{#1}%
\providecommand \citenamefont [1]{#1}%
\providecommand \href@noop [0]{\@secondoftwo}%
\providecommand \href [0]{\begingroup \@sanitize@url \@href}%
\providecommand \@href[1]{\@@startlink{#1}\@@href}%
\providecommand \@@href[1]{\endgroup#1\@@endlink}%
\providecommand \@sanitize@url [0]{\catcode `\\12\catcode `\$12\catcode
  `\&12\catcode `\#12\catcode `\^12\catcode `\_12\catcode `\%12\relax}%
\providecommand \@@startlink[1]{}%
\providecommand \@@endlink[0]{}%
\providecommand \url  [0]{\begingroup\@sanitize@url \@url }%
\providecommand \@url [1]{\endgroup\@href {#1}{\urlprefix }}%
\providecommand \urlprefix  [0]{URL }%
\providecommand \Eprint [0]{\href }%
\providecommand \doibase [0]{http://dx.doi.org/}%
\providecommand \selectlanguage [0]{\@gobble}%
\providecommand \bibinfo  [0]{\@secondoftwo}%
\providecommand \bibfield  [0]{\@secondoftwo}%
\providecommand \translation [1]{[#1]}%
\providecommand \BibitemOpen [0]{}%
\providecommand \bibitemStop [0]{}%
\providecommand \bibitemNoStop [0]{.\EOS\space}%
\providecommand \EOS [0]{\spacefactor3000\relax}%
\providecommand \BibitemShut  [1]{\csname bibitem#1\endcsname}%
\let\auto@bib@innerbib\@empty
%</preamble>

\bibitem [{\citenamefont {Dagotto}\ and\ \citenamefont
  {Tokura}(2008)}]{Dagotto2008}%
  %\BibitemOpen
  \bibfield  {author} {\bibinfo {author} {\bibfnamefont {E.}~\bibnamefont
  {Dagotto}}\ and\ \bibinfo {author} {\bibfnamefont {Y.}~\bibnamefont
  {Tokura}},\ }\href {\doibase 10.1557/mrs2008.223} {\bibfield  {journal}
  {\bibinfo  {journal} {MRS Bull.}\ }\textbf {\bibinfo {volume} {33}},\
  \bibinfo {pages} {1037} (\bibinfo {year} {2008})}.
  %\BibitemShut {NoStop}%

\bibitem [{\citenamefont {Zubko}\ \emph {et~al.}(2011)\citenamefont {Zubko},
  \citenamefont {Gariglio}, \citenamefont {Gabay}, \citenamefont {Ghosez},\
  and\ \citenamefont {Triscone}}]{Zubko2011}%
  %\BibitemOpen
  \bibfield  {author} {\bibinfo {author} {\bibfnamefont {P.}~\bibnamefont
  {Zubko}}, \bibinfo {author} {\bibfnamefont {S.}~\bibnamefont {Gariglio}},
  \bibinfo {author} {\bibfnamefont {M.}~\bibnamefont {Gabay}}, \bibinfo
  {author} {\bibfnamefont {P.}~\bibnamefont {Ghosez}}, \ and\ \bibinfo {author}
  {\bibfnamefont {J.-M.}\ \bibnamefont {Triscone}},\ }\href {\doibase
  10.1146/annurev-conmatphys-062910-140445} {\bibfield  {journal} {\bibinfo
  {journal} {Annu. Rev. Condens. Matter Phys.}\ }\textbf {\bibinfo {volume}
  {2}},\ \bibinfo {pages} {141} (\bibinfo {year} {2011})}.
  %\BibitemShut {NoStop}%

\bibitem [{\citenamefont {Goldschmidt}(1926)}]{Goldschmidt1926}%
  %\BibitemOpen
  \bibfield  {author} {\bibinfo {author} {\bibfnamefont {V.~M.}\ \bibnamefont
  {Goldschmidt}},\ }\href
  {http://www.springerlink.com/index/UV03N13397203255.pdf} {\bibfield
  {journal} {\bibinfo  {journal} {Naturwissenschaften}\ }\textbf {\bibinfo
  {volume} {14}},\ \bibinfo {pages} {477} (\bibinfo {year} {1926})}.
  %\BibitemShut{NoStop}%

\bibitem [{\citenamefont {Medarde}(1997)}]{Medarde1997}%
  %\BibitemOpen
  \bibfield  {author} {\bibinfo {author} {\bibfnamefont {M.~L.}\ \bibnamefont
  {Medarde}},\ }\href {http://iopscience.iop.org/0953-8984/9/8/003} {\bibfield
  {journal} {\bibinfo  {journal} {J. Phys. Condens. Matter}\ }\textbf {\bibinfo
  {volume} {9}},\ \bibinfo {pages} {1679} (\bibinfo {year} {1997})}.
  %\BibitemShut{NoStop}%

\bibitem [{\citenamefont {Kimura}\ \emph {et~al.}(2003)\citenamefont {Kimura},
  \citenamefont {Ishihara}, \citenamefont {Shintani}, \citenamefont {Arima},
  \citenamefont {Takahashi}, \citenamefont {Ishizaka},\ and\ \citenamefont
  {Tokura}}]{Kimura2003}%
  %\BibitemOpen
  \bibfield  {author} {\bibinfo {author} {\bibfnamefont {T.}~\bibnamefont
  {Kimura}}, \bibinfo {author} {\bibfnamefont {S.}~\bibnamefont {Ishihara}},
  \bibinfo {author} {\bibfnamefont {H.}~\bibnamefont {Shintani}}, \bibinfo
  {author} {\bibfnamefont {T.}~\bibnamefont {Arima}}, \bibinfo {author}
  {\bibfnamefont {K.}~\bibnamefont {Takahashi}}, \bibinfo {author}
  {\bibfnamefont {K.}~\bibnamefont {Ishizaka}}, \ and\ \bibinfo {author}
  {\bibfnamefont {Y.}~\bibnamefont {Tokura}},\ }\href {\doibase
  10.1103/PhysRevB.68.060403} {\bibfield  {journal} {\bibinfo  {journal} {Phys.
  Rev. B}\ }\textbf {\bibinfo {volume} {68}},\ \bibinfo {pages} {060403}
  (\bibinfo {year} {2003})}.
  %\BibitemShut {NoStop}%

\bibitem [{\citenamefont {Ohtomo}\ and\ \citenamefont
  {Hwang}(2004)}]{Ohtomo2004}%
  %\BibitemOpen
  \bibfield  {author} {\bibinfo {author} {\bibfnamefont {A.}~\bibnamefont
  {Ohtomo}}\ and\ \bibinfo {author} {\bibfnamefont {H.~Y.}\ \bibnamefont
  {Hwang}},\ }\href {\doibase 10.1038/nature02308} {\bibfield  {journal}
  {\bibinfo  {journal} {Nature}\ }\textbf {\bibinfo {volume} {427}},\ \bibinfo
  {pages} {423} (\bibinfo {year} {2004})}.
  %\BibitemShut {NoStop}%

\bibitem [{\citenamefont {Dikin}\ \emph {et~al.}(2011)\citenamefont {Dikin},
  \citenamefont {Mehta}, \citenamefont {Bark}, \citenamefont {Folkman},
  \citenamefont {Eom},\ and\ \citenamefont {Chandrasekhar}}]{Dikin2011}%
  %\BibitemOpen
  \bibfield  {author} {\bibinfo {author} {\bibfnamefont {D.~A.}\ \bibnamefont
  {Dikin}}, \bibinfo {author} {\bibfnamefont {M.}~\bibnamefont {Mehta}},
  \bibinfo {author} {\bibfnamefont {C.~W.}\ \bibnamefont {Bark}}, \bibinfo
  {author} {\bibfnamefont {C.~M.}\ \bibnamefont {Folkman}}, \bibinfo {author}
  {\bibfnamefont {C.~B.}\ \bibnamefont {Eom}}, \ and\ \bibinfo {author}
  {\bibfnamefont {V.}~\bibnamefont {Chandrasekhar}},\ }\href {\doibase
  10.1103/PhysRevLett.107.056802} {\bibfield  {journal} {\bibinfo  {journal}
  {Phys. Rev. Lett.}\ }\textbf {\bibinfo {volume} {107}},\ \bibinfo {pages}
  {056802} (\bibinfo {year} {2011})}.
  %\BibitemShut {NoStop}%

\bibitem [{\citenamefont {Li}\ \emph {et~al.}(2011)\citenamefont {Li},
  \citenamefont {Richter}, \citenamefont {Mannhart},\ and\ \citenamefont
  {Ashoori}}]{Li2011a}%
  %\BibitemOpen
  \bibfield  {author} {\bibinfo {author} {\bibfnamefont {L.}~\bibnamefont
  {Li}}, \bibinfo {author} {\bibfnamefont {C.}~\bibnamefont {Richter}},
  \bibinfo {author} {\bibfnamefont {J.}~\bibnamefont {Mannhart}}, \ and\
  \bibinfo {author} {\bibfnamefont {R.~C.}\ \bibnamefont {Ashoori}},\ }\href
  {\doibase 10.1038/nphys2080} {\bibfield  {journal} {\bibinfo  {journal} {Nat.
  Phys.}\ }\textbf {\bibinfo {volume} {7}},\ \bibinfo {pages} {762} (\bibinfo
  {year} {2011})}.
  %\BibitemShut {NoStop}%

\bibitem [{\citenamefont {Cheng}\ \emph {et~al.}(2015)\citenamefont {Cheng},
  \citenamefont {Tomczyk}, \citenamefont {Lu}, \citenamefont {Veazey},
  \citenamefont {Huang}, \citenamefont {Irvin}, \citenamefont {Ryu},
  \citenamefont {Lee}, \citenamefont {Eom}, \citenamefont {Hellberg},\ and\
  \citenamefont {Levy}}]{Cheng2015}%
  %\BibitemOpen
  \bibfield  {author} {\bibinfo {author} {\bibfnamefont {G.}~\bibnamefont
  {Cheng}}, \bibinfo {author} {\bibfnamefont {M.}~\bibnamefont {Tomczyk}},
  \bibinfo {author} {\bibfnamefont {S.}~\bibnamefont {Lu}}, \bibinfo {author}
  {\bibfnamefont {J.~P.}\ \bibnamefont {Veazey}}, \bibinfo {author}
  {\bibfnamefont {M.}~\bibnamefont {Huang}}, \bibinfo {author} {\bibfnamefont
  {P.}~\bibnamefont {Irvin}}, \bibinfo {author} {\bibfnamefont
  {S.}~\bibnamefont {Ryu}}, \bibinfo {author} {\bibfnamefont {H.}~\bibnamefont
  {Lee}}, \bibinfo {author} {\bibfnamefont {C.-B.}\ \bibnamefont {Eom}},
  \bibinfo {author} {\bibfnamefont {C.~S.}\ \bibnamefont {Hellberg}}, \ and\
  \bibinfo {author} {\bibfnamefont {J.}~\bibnamefont {Levy}},\ }\href {\doibase
  10.1038/nature14398} {\bibfield  {journal} {\bibinfo  {journal} {Nature}\
  }\textbf {\bibinfo {volume} {521}},\ \bibinfo {pages} {196} (\bibinfo {year}
  {2015})}.
  %\BibitemShut {NoStop}%

\bibitem [{\citenamefont {Santander-Syro}\ \emph {et~al.}(2011)\citenamefont
  {Santander-Syro}, \citenamefont {Copie}, \citenamefont {Kondo}, \citenamefont
  {Fortuna}, \citenamefont {Pailh{\`{e}}s}, \citenamefont {Weht}, \citenamefont
  {Qiu}, \citenamefont {Bertran}, \citenamefont {Nicolaou}, \citenamefont
  {Taleb-Ibrahimi}, \citenamefont {{Le F{\`{e}}vre}}, \citenamefont {Herranz},
  \citenamefont {Bibes}, \citenamefont {Reyren}, \citenamefont {Apertet},
  \citenamefont {Lecoeur}, \citenamefont {Barth{\'{e}}l{\'{e}}my},\ and\
  \citenamefont {Rozenberg}}]{Santander-Syro2011}%
  %\BibitemOpen
  \bibfield  {author} {\bibinfo {author} {\bibfnamefont {A.~F.}\ \bibnamefont
  {Santander-Syro}}, \bibinfo {author} {\bibfnamefont {O.}~\bibnamefont
  {Copie}}, \bibinfo {author} {\bibfnamefont {T.}~\bibnamefont {Kondo}},
  \bibinfo {author} {\bibfnamefont {F.}~\bibnamefont {Fortuna}}, \bibinfo
  {author} {\bibfnamefont {S.}~\bibnamefont {Pailh{\`{e}}s}}, \bibinfo {author}
  {\bibfnamefont {R.}~\bibnamefont {Weht}}, \bibinfo {author} {\bibfnamefont
  {X.~G.}\ \bibnamefont {Qiu}}, \bibinfo {author} {\bibfnamefont
  {F.}~\bibnamefont {Bertran}}, \bibinfo {author} {\bibfnamefont
  {A.}~\bibnamefont {Nicolaou}}, \bibinfo {author} {\bibfnamefont
  {A.}~\bibnamefont {Taleb-Ibrahimi}}, \bibinfo {author} {\bibfnamefont
  {P.}~\bibnamefont {{Le F{\`{e}}vre}}}, \bibinfo {author} {\bibfnamefont
  {G.}~\bibnamefont {Herranz}}, \bibinfo {author} {\bibfnamefont
  {M.}~\bibnamefont {Bibes}}, \bibinfo {author} {\bibfnamefont
  {N.}~\bibnamefont {Reyren}}, \bibinfo {author} {\bibfnamefont
  {Y.}~\bibnamefont {Apertet}}, \bibinfo {author} {\bibfnamefont
  {P.}~\bibnamefont {Lecoeur}}, \bibinfo {author} {\bibfnamefont
  {A.}~\bibnamefont {Barth{\'{e}}l{\'{e}}my}}, \ and\ \bibinfo {author}
  {\bibfnamefont {M.~J.}\ \bibnamefont {Rozenberg}},\ }\href {\doibase
  10.1038/nature09720} {\bibfield  {journal} {\bibinfo  {journal} {Nature}\
  }\textbf {\bibinfo {volume} {469}},\ \bibinfo {pages} {189} (\bibinfo {year}
  {2011})}.
  %\BibitemShut {NoStop}%

\bibitem [{\citenamefont {Meevasana}\ \emph {et~al.}(2011)\citenamefont
  {Meevasana}, \citenamefont {King}, \citenamefont {He}, \citenamefont {Mo},
  \citenamefont {Hashimoto}, \citenamefont {Tamai}, \citenamefont
  {Songsiriritthigul}, \citenamefont {Baumberger},\ and\ \citenamefont
  {Shen}}]{Meevasana2011}%
  %\BibitemOpen
  \bibfield  {author} {\bibinfo {author} {\bibfnamefont {W.}~\bibnamefont
  {Meevasana}}, \bibinfo {author} {\bibfnamefont {P.~D.~C.}\ \bibnamefont
  {King}}, \bibinfo {author} {\bibfnamefont {R.~H.}\ \bibnamefont {He}},
  \bibinfo {author} {\bibfnamefont {S.-K.}\ \bibnamefont {Mo}}, \bibinfo
  {author} {\bibfnamefont {M.}~\bibnamefont {Hashimoto}}, \bibinfo {author}
  {\bibfnamefont {A.}~\bibnamefont {Tamai}}, \bibinfo {author} {\bibfnamefont
  {P.}~\bibnamefont {Songsiriritthigul}}, \bibinfo {author} {\bibfnamefont
  {F.}~\bibnamefont {Baumberger}}, \ and\ \bibinfo {author} {\bibfnamefont
  {Z.-X.}\ \bibnamefont {Shen}},\ }\href {\doibase 10.1038/NMAT2943} {\bibfield
   {journal} {\bibinfo  {journal} {Nat. Mater.}\ }\textbf {\bibinfo {volume}
  {10}},\ \bibinfo {pages} {114} (\bibinfo {year} {2011})}.
  %\BibitemShut{NoStop}%

\bibitem [{\citenamefont {Wang}\ \emph {et~al.}(2014)\citenamefont {Wang},
  \citenamefont {Zhong}, \citenamefont {Hao}, \citenamefont {Gerhold},
  \citenamefont {Stoger}, \citenamefont {Schmid}, \citenamefont
  {Sanchez-Barriga}, \citenamefont {Varykhalov}, \citenamefont {Franchini},
  \citenamefont {Held},\ and\ \citenamefont {Diebold}}]{Wang2014}%
  %\BibitemOpen
  \bibfield  {author} {\bibinfo {author} {\bibfnamefont {Z.}~\bibnamefont
  {Wang}}, \bibinfo {author} {\bibfnamefont {Z.}~\bibnamefont {Zhong}},
  \bibinfo {author} {\bibfnamefont {X.}~\bibnamefont {Hao}}, \bibinfo {author}
  {\bibfnamefont {S.}~\bibnamefont {Gerhold}}, \bibinfo {author} {\bibfnamefont
  {B.}~\bibnamefont {Stoger}}, \bibinfo {author} {\bibfnamefont
  {M.}~\bibnamefont {Schmid}}, \bibinfo {author} {\bibfnamefont
  {J.}~\bibnamefont {Sanchez-Barriga}}, \bibinfo {author} {\bibfnamefont
  {A.}~\bibnamefont {Varykhalov}}, \bibinfo {author} {\bibfnamefont
  {C.}~\bibnamefont {Franchini}}, \bibinfo {author} {\bibfnamefont
  {K.}~\bibnamefont {Held}}, \ and\ \bibinfo {author} {\bibfnamefont
  {U.}~\bibnamefont {Diebold}},\ }\href {\doibase 10.1073/pnas.1318304111}
  {\bibfield  {journal} {\bibinfo  {journal} {Proc. Natl. Acad. Sci.}\ }\textbf
  {\bibinfo {volume} {111}},\ \bibinfo {pages} {3933} (\bibinfo {year}
  {2014})},\ \Eprint {http://arxiv.org/abs/arXiv:1309.7042v1}
  {arXiv:arXiv:1309.7042v1}. 
  %\BibitemShut {NoStop}%

\bibitem [{\citenamefont {R{\"{o}}del}\ \emph {et~al.}(2014)\citenamefont
  {R{\"{o}}del}, \citenamefont {Bareille}, \citenamefont {Fortuna},
  \citenamefont {Baumier}, \citenamefont {Bertran}, \citenamefont {{Le
  F{\`{e}}vre}}, \citenamefont {Gabay}, \citenamefont {Cubelos}, \citenamefont
  {Rozenberg}, \citenamefont {Maroutian}, \citenamefont {Lecoeur},\ and\
  \citenamefont {Santander-Syro}}]{Roedel2014}%
  %\BibitemOpen
  \bibfield  {author} {\bibinfo {author} {\bibfnamefont {T.~C.}\ \bibnamefont
  {R{\"{o}}del}}, \bibinfo {author} {\bibfnamefont {C.}~\bibnamefont
  {Bareille}}, \bibinfo {author} {\bibfnamefont {F.}~\bibnamefont {Fortuna}},
  \bibinfo {author} {\bibfnamefont {C.}~\bibnamefont {Baumier}}, \bibinfo
  {author} {\bibfnamefont {F.}~\bibnamefont {Bertran}}, \bibinfo {author}
  {\bibfnamefont {P.}~\bibnamefont {{Le F{\`{e}}vre}}}, \bibinfo {author}
  {\bibfnamefont {M.}~\bibnamefont {Gabay}}, \bibinfo {author} {\bibfnamefont
  {O.~H.}\ \bibnamefont {Cubelos}}, \bibinfo {author} {\bibfnamefont {M.~J.}\
  \bibnamefont {Rozenberg}}, \bibinfo {author} {\bibfnamefont {T.}~\bibnamefont
  {Maroutian}}, \bibinfo {author} {\bibfnamefont {P.}~\bibnamefont {Lecoeur}},
  \ and\ \bibinfo {author} {\bibfnamefont {A.~F.}\ \bibnamefont
  {Santander-Syro}},\ }\href {\doibase 10.1103/PhysRevApplied.1.051002}
  {\bibfield  {journal} {\bibinfo  {journal} {Phys. Rev. Appl.}\ }\textbf
  {\bibinfo {volume} {1}},\ \bibinfo {pages} {051002} (\bibinfo {year}
  {2014})}.
  %\BibitemShut {NoStop}%

\bibitem [{\citenamefont {{McKeown Walker}}\ \emph {et~al.}(2014)\citenamefont
  {{McKeown Walker}}, \citenamefont {de~la Torre}, \citenamefont {Bruno},
  \citenamefont {Tamai}, \citenamefont {Kim}, \citenamefont {Hoesch},
  \citenamefont {Shi}, \citenamefont {Bahramy}, \citenamefont {King},\ and\
  \citenamefont {Baumberger}}]{Walker2014}%
  %\BibitemOpen
  \bibfield  {author} {\bibinfo {author} {\bibfnamefont {S.}~\bibnamefont
  {{McKeown Walker}}}, \bibinfo {author} {\bibfnamefont {A.}~\bibnamefont
  {de~la Torre}}, \bibinfo {author} {\bibfnamefont {F.~Y.}\ \bibnamefont
  {Bruno}}, \bibinfo {author} {\bibfnamefont {A.}~\bibnamefont {Tamai}},
  \bibinfo {author} {\bibfnamefont {T.~K.}\ \bibnamefont {Kim}}, \bibinfo
  {author} {\bibfnamefont {M.}~\bibnamefont {Hoesch}}, \bibinfo {author}
  {\bibfnamefont {M.}~\bibnamefont {Shi}}, \bibinfo {author} {\bibfnamefont
  {M.~S.}\ \bibnamefont {Bahramy}}, \bibinfo {author} {\bibfnamefont
  {P.~D.~C.}\ \bibnamefont {King}}, \ and\ \bibinfo {author} {\bibfnamefont
  {F.}~\bibnamefont {Baumberger}},\ }\href {\doibase
  10.1103/PhysRevLett.113.177601} {\bibfield  {journal} {\bibinfo  {journal}
  {Phys. Rev. Lett.}\ }\textbf {\bibinfo {volume} {113}},\ \bibinfo {pages}
  {177601} (\bibinfo {year} {2014})}.
  %\BibitemShut {NoStop}%

\bibitem [{\citenamefont {King}\ \emph {et~al.}(2012)\citenamefont {King},
  \citenamefont {He}, \citenamefont {Eknapakul}, \citenamefont {Buaphet},
  \citenamefont {Mo}, \citenamefont {Kaneko}, \citenamefont {Harashima},
  \citenamefont {Hikita}, \citenamefont {Bahramy}, \citenamefont {Bell},
  \citenamefont {Hussain}, \citenamefont {Tokura}, \citenamefont {Shen},
  \citenamefont {Hwang}, \citenamefont {Baumberger},\ and\ \citenamefont
  {Meevasana}}]{King2012}%
  %\BibitemOpen
  \bibfield  {author} {\bibinfo {author} {\bibfnamefont {P.~D.~C.}\
  \bibnamefont {King}}, \bibinfo {author} {\bibfnamefont {R.~H.}\ \bibnamefont
  {He}}, \bibinfo {author} {\bibfnamefont {T.}~\bibnamefont {Eknapakul}},
  \bibinfo {author} {\bibfnamefont {P.}~\bibnamefont {Buaphet}}, \bibinfo
  {author} {\bibfnamefont {S.-K.}\ \bibnamefont {Mo}}, \bibinfo {author}
  {\bibfnamefont {Y.}~\bibnamefont {Kaneko}}, \bibinfo {author} {\bibfnamefont
  {S.}~\bibnamefont {Harashima}}, \bibinfo {author} {\bibfnamefont
  {Y.}~\bibnamefont {Hikita}}, \bibinfo {author} {\bibfnamefont {M.~S.}\
  \bibnamefont {Bahramy}}, \bibinfo {author} {\bibfnamefont {C.}~\bibnamefont
  {Bell}}, \bibinfo {author} {\bibfnamefont {Z.}~\bibnamefont {Hussain}},
  \bibinfo {author} {\bibfnamefont {Y.}~\bibnamefont {Tokura}}, \bibinfo
  {author} {\bibfnamefont {Z.-X.}\ \bibnamefont {Shen}}, \bibinfo {author}
  {\bibfnamefont {H.~Y.}\ \bibnamefont {Hwang}}, \bibinfo {author}
  {\bibfnamefont {F.}~\bibnamefont {Baumberger}}, \ and\ \bibinfo {author}
  {\bibfnamefont {W.}~\bibnamefont {Meevasana}},\ }\href {\doibase
  10.1103/PhysRevLett.108.117602} {\bibfield  {journal} {\bibinfo  {journal}
  {Phys. Rev. Lett.}\ }\textbf {\bibinfo {volume} {108}},\ \bibinfo {pages}
  {117602} (\bibinfo {year} {2012})}.
  %\BibitemShut {NoStop}%

\bibitem [{\citenamefont {Santander-Syro}\ \emph {et~al.}(2012)\citenamefont
  {Santander-Syro}, \citenamefont {Bareille}, \citenamefont {Fortuna},
  \citenamefont {Copie}, \citenamefont {Gabay}, \citenamefont {Bertran},
  \citenamefont {Taleb-Ibrahimi}, \citenamefont {{Le F{\`{e}}vre}},
  \citenamefont {Herranz}, \citenamefont {Reyren}, \citenamefont {Bibes},
  \citenamefont {Barth{\'{e}}l{\'{e}}my}, \citenamefont {Lecoeur},
  \citenamefont {Guevara},\ and\ \citenamefont
  {Rozenberg}}]{Santander-Syro2012}%
  %\BibitemOpen
  \bibfield  {author} {\bibinfo {author} {\bibfnamefont {A.~F.}\ \bibnamefont
  {Santander-Syro}}, \bibinfo {author} {\bibfnamefont {C.}~\bibnamefont
  {Bareille}}, \bibinfo {author} {\bibfnamefont {F.}~\bibnamefont {Fortuna}},
  \bibinfo {author} {\bibfnamefont {O.}~\bibnamefont {Copie}}, \bibinfo
  {author} {\bibfnamefont {M.}~\bibnamefont {Gabay}}, \bibinfo {author}
  {\bibfnamefont {F.}~\bibnamefont {Bertran}}, \bibinfo {author} {\bibfnamefont
  {A.}~\bibnamefont {Taleb-Ibrahimi}}, \bibinfo {author} {\bibfnamefont
  {P.}~\bibnamefont {{Le F{\`{e}}vre}}}, \bibinfo {author} {\bibfnamefont
  {G.}~\bibnamefont {Herranz}}, \bibinfo {author} {\bibfnamefont
  {N.}~\bibnamefont {Reyren}}, \bibinfo {author} {\bibfnamefont
  {M.}~\bibnamefont {Bibes}}, \bibinfo {author} {\bibfnamefont
  {A.}~\bibnamefont {Barth{\'{e}}l{\'{e}}my}}, \bibinfo {author} {\bibfnamefont
  {P.}~\bibnamefont {Lecoeur}}, \bibinfo {author} {\bibfnamefont
  {J.}~\bibnamefont {Guevara}}, \ and\ \bibinfo {author} {\bibfnamefont
  {M.~J.}\ \bibnamefont {Rozenberg}},\ }\href {\doibase
  10.1103/PhysRevB.86.121107} {\bibfield  {journal} {\bibinfo  {journal} {Phys.
  Rev. B}\ }\textbf {\bibinfo {volume} {86}},\ \bibinfo {pages} {121107(R)}
  (\bibinfo {year} {2012})}.
  %\BibitemShut {NoStop}%

\bibitem [{\citenamefont {Bareille}\ \emph {et~al.}(2014)\citenamefont
  {Bareille}, \citenamefont {Fortuna}, \citenamefont {R{\"{o}}del},
  \citenamefont {Bertran}, \citenamefont {Gabay}, \citenamefont {Cubelos},
  \citenamefont {Taleb-Ibrahimi}, \citenamefont {{Le F{\`{e}}vre}},
  \citenamefont {Bibes}, \citenamefont {Barth{\'{e}}l{\'{e}}my}, \citenamefont
  {Maroutian}, \citenamefont {Lecoeur}, \citenamefont {Rozenberg},\ and\
  \citenamefont {Santander-Syro}}]{Bareille2014}%
  %\BibitemOpen
  \bibfield  {author} {\bibinfo {author} {\bibfnamefont {C.}~\bibnamefont
  {Bareille}}, \bibinfo {author} {\bibfnamefont {F.}~\bibnamefont {Fortuna}},
  \bibinfo {author} {\bibfnamefont {T.~C.}\ \bibnamefont {R{\"{o}}del}},
  \bibinfo {author} {\bibfnamefont {F.}~\bibnamefont {Bertran}}, \bibinfo
  {author} {\bibfnamefont {M.}~\bibnamefont {Gabay}}, \bibinfo {author}
  {\bibfnamefont {O.~H.}\ \bibnamefont {Cubelos}}, \bibinfo {author}
  {\bibfnamefont {A.}~\bibnamefont {Taleb-Ibrahimi}}, \bibinfo {author}
  {\bibfnamefont {P.}~\bibnamefont {{Le F{\`{e}}vre}}}, \bibinfo {author}
  {\bibfnamefont {M.}~\bibnamefont {Bibes}}, \bibinfo {author} {\bibfnamefont
  {A.}~\bibnamefont {Barth{\'{e}}l{\'{e}}my}}, \bibinfo {author} {\bibfnamefont
  {T.}~\bibnamefont {Maroutian}}, \bibinfo {author} {\bibfnamefont
  {P.}~\bibnamefont {Lecoeur}}, \bibinfo {author} {\bibfnamefont {M.~J.}\
  \bibnamefont {Rozenberg}}, \ and\ \bibinfo {author} {\bibfnamefont {A.~F.}\
  \bibnamefont {Santander-Syro}},\ }\href {\doibase 10.1038/srep03586}
  {\bibfield  {journal} {\bibinfo  {journal} {Sci. Rep.}\ }\textbf {\bibinfo
  {volume} {4}},\ \bibinfo {pages} {3586} (\bibinfo {year} {2014})}.
  %\BibitemShut{NoStop}%

\bibitem [{\citenamefont {R{\"{o}}del}\ \emph {et~al.}(2015)\citenamefont
  {R{\"{o}}del}, \citenamefont {Fortuna}, \citenamefont {Bertran},
  \citenamefont {Gabay}, \citenamefont {Rozenberg}, \citenamefont
  {Santander-Syro},\ and\ \citenamefont {F{\`{e}}vre}}]{Roedel2015}%
  %\BibitemOpen
  \bibfield  {author} {\bibinfo {author} {\bibfnamefont {T.~C.}\ \bibnamefont
  {R{\"{o}}del}}, \bibinfo {author} {\bibfnamefont {F.}~\bibnamefont
  {Fortuna}}, \bibinfo {author} {\bibfnamefont {F.}~\bibnamefont {Bertran}},
  \bibinfo {author} {\bibfnamefont {M.}~\bibnamefont {Gabay}}, \bibinfo
  {author} {\bibfnamefont {M.~J.}\ \bibnamefont {Rozenberg}}, \bibinfo {author}
  {\bibfnamefont {a.~F.}\ \bibnamefont {Santander-Syro}}, \ and\ \bibinfo
  {author} {\bibfnamefont {P.~L.}\ \bibnamefont {F{\`{e}}vre}},\ }\href
  {\doibase 10.1103/PhysRevB.92.041106} {\bibfield  {journal} {\bibinfo
  {journal} {Phys. Rev. B}\ }\textbf {\bibinfo {volume} {92}},\ \bibinfo
  {pages} {041106(R)} (\bibinfo {year} {2015})},\ \Eprint
  {http://arxiv.org/abs/1507.03916} {arXiv:1507.03916}. 
  %\BibitemShut {NoStop}%

\bibitem [{\citenamefont {R{\"{o}}del}\ \emph {et~al.}(2016)\citenamefont
  {R{\"{o}}del}, \citenamefont {Fortuna}, \citenamefont {Sengupta},
  \citenamefont {Frantzeskakis}, \citenamefont {{Le F{\`{e}}vre}},
  \citenamefont {Bertran}, \citenamefont {Mercey}, \citenamefont {Matzen},
  \citenamefont {Agnus}, \citenamefont {Maroutian}, \citenamefont {Lecoeur},\
  and\ \citenamefont {Santander-Syro}}]{Rodel2016}%
  %\BibitemOpen
  \bibfield  {author} {\bibinfo {author} {\bibfnamefont {T.~C.}\ \bibnamefont
  {R{\"{o}}del}}, \bibinfo {author} {\bibfnamefont {F.}~\bibnamefont
  {Fortuna}}, \bibinfo {author} {\bibfnamefont {S.}~\bibnamefont {Sengupta}},
  \bibinfo {author} {\bibfnamefont {E.}~\bibnamefont {Frantzeskakis}}, \bibinfo
  {author} {\bibfnamefont {P.}~\bibnamefont {{Le F{\`{e}}vre}}}, \bibinfo
  {author} {\bibfnamefont {F.}~\bibnamefont {Bertran}}, \bibinfo {author}
  {\bibfnamefont {B.}~\bibnamefont {Mercey}}, \bibinfo {author} {\bibfnamefont
  {S.}~\bibnamefont {Matzen}}, \bibinfo {author} {\bibfnamefont
  {G.}~\bibnamefont {Agnus}}, \bibinfo {author} {\bibfnamefont
  {T.}~\bibnamefont {Maroutian}}, \bibinfo {author} {\bibfnamefont
  {P.}~\bibnamefont {Lecoeur}}, \ and\ \bibinfo {author} {\bibfnamefont
  {A.~F.}\ \bibnamefont {Santander-Syro}},\ }\href {\doibase
  10.1002/adma.201505021} {\bibfield  {journal} {\bibinfo  {journal} {Adv.
  Mater.}\ }\textbf {\bibinfo {volume} {28}},\ \bibinfo {pages} {1976}
  (\bibinfo {year} {2016})}.
  %\BibitemShut {NoStop}%

\bibitem [{\citenamefont {Chen}\ \emph {et~al.}(2015)\citenamefont {Chen},
  \citenamefont {Avila}, \citenamefont {Frantzeskakis}, \citenamefont {Levy},\
  and\ \citenamefont {Asensio}}]{Chen2015e}%
  %\BibitemOpen
  \bibfield  {author} {\bibinfo {author} {\bibfnamefont {C.}~\bibnamefont
  {Chen}}, \bibinfo {author} {\bibfnamefont {J.}~\bibnamefont {Avila}},
  \bibinfo {author} {\bibfnamefont {E.}~\bibnamefont {Frantzeskakis}}, \bibinfo
  {author} {\bibfnamefont {A.}~\bibnamefont {Levy}}, \ and\ \bibinfo {author}
  {\bibfnamefont {M.~C.}\ \bibnamefont {Asensio}},\ }\href {\doibase
  10.1038/ncomms9585} {\bibfield  {journal} {\bibinfo  {journal} {Nat.
  Commun.}\ }\textbf {\bibinfo {volume} {6}},\ \bibinfo {pages} {8585}
  (\bibinfo {year} {2015})}.
  %\BibitemShut {NoStop}%

\bibitem [{\citenamefont {Wang}\ \emph {et~al.}(2016)\citenamefont {Wang},
  \citenamefont {{McKeown Walker}}, \citenamefont {Tamai}, \citenamefont
  {Ristic}, \citenamefont {Bruno}, \citenamefont {de~la Torre}, \citenamefont
  {Ricc{\`{o}}}, \citenamefont {Plumb}, \citenamefont {Shi}, \citenamefont
  {Hlawenka}, \citenamefont {Sanchez-Barriga}, \citenamefont {Varykhalov},
  \citenamefont {Kim}, \citenamefont {Hoesch}, \citenamefont {King},
  \citenamefont {Meevasana}, \citenamefont {Diebold}, \citenamefont {Mesot},
  \citenamefont {Radovi\'{c}},\ and\ \citenamefont {Baumberger}}]{Wang2015a}%
  %\BibitemOpen
  \bibfield  {author} {\bibinfo {author} {\bibfnamefont {Z.}~\bibnamefont
  {Wang}}, \bibinfo {author} {\bibfnamefont {S.}~\bibnamefont {{McKeown
  Walker}}}, \bibinfo {author} {\bibfnamefont {A.}~\bibnamefont {Tamai}},
  \bibinfo {author} {\bibfnamefont {Z.}~\bibnamefont {Ristic}}, \bibinfo
  {author} {\bibfnamefont {F.~Y.}\ \bibnamefont {Bruno}}, \bibinfo {author}
  {\bibfnamefont {A.}~\bibnamefont {de~la Torre}}, \bibinfo {author}
  {\bibfnamefont {S.}~\bibnamefont {Ricc{\`{o}}}}, \bibinfo {author}
  {\bibfnamefont {N.}~\bibnamefont {Plumb}}, \bibinfo {author} {\bibfnamefont
  {M.}~\bibnamefont {Shi}}, \bibinfo {author} {\bibfnamefont {P.}~\bibnamefont
  {Hlawenka}}, \bibinfo {author} {\bibfnamefont {J.}~\bibnamefont
  {Sanchez-Barriga}}, \bibinfo {author} {\bibfnamefont {A.}~\bibnamefont
  {Varykhalov}}, \bibinfo {author} {\bibfnamefont {T.~K.}\ \bibnamefont {Kim}},
  \bibinfo {author} {\bibfnamefont {M.}~\bibnamefont {Hoesch}}, \bibinfo
  {author} {\bibfnamefont {P.~D.~C.}\ \bibnamefont {King}}, \bibinfo {author}
  {\bibfnamefont {W.}~\bibnamefont {Meevasana}}, \bibinfo {author}
  {\bibfnamefont {U.}~\bibnamefont {Diebold}}, \bibinfo {author} {\bibfnamefont
  {J.}~\bibnamefont {Mesot}}, \bibinfo {author} {\bibfnamefont
  {M.}~\bibnamefont {Radovi\'{c}}}, \ and\ \bibinfo {author} {\bibfnamefont
  {F.}~\bibnamefont {Baumberger}},\ }\href@noop {} {\bibfield  {journal}
  {\bibinfo  {journal} {Nat. Mater.}\ ,\ \bibinfo {pages}
  {doi:10.1038/nmat4623}} (\bibinfo {year} {2016})},\ \Eprint
  {http://arxiv.org/abs/arXiv:1506.01191v1} {arXiv:arXiv:1506.01191v1}.
  %\BibitemShut {NoStop}%

\bibitem [{\citenamefont {Rondinelli}\ \emph {et~al.}(2012)\citenamefont
  {Rondinelli}, \citenamefont {May},\ and\ \citenamefont
  {Freeland}}]{Rondinelli2012a}%
  %\BibitemOpen
  \bibfield  {author} {\bibinfo {author} {\bibfnamefont {J.~M.}\ \bibnamefont
  {Rondinelli}}, \bibinfo {author} {\bibfnamefont {S.~J.}\ \bibnamefont {May}},
  \ and\ \bibinfo {author} {\bibfnamefont {J.~W.}\ \bibnamefont {Freeland}},\
  }\href {\doibase 10.1557/mrs.2012.49} {\bibfield  {journal} {\bibinfo
  {journal} {MRS Bull.}\ }\textbf {\bibinfo {volume} {37}},\ \bibinfo {pages}
  {261} (\bibinfo {year} {2012})}.
  %\BibitemShut {NoStop}%

\bibitem [{\citenamefont {Ganguli}\ and\ \citenamefont
  {Kelly}(2014)}]{Ganguli2014}%
  %\BibitemOpen
  \bibfield  {author} {\bibinfo {author} {\bibfnamefont {N.}~\bibnamefont
  {Ganguli}}\ and\ \bibinfo {author} {\bibfnamefont {P.~J.}\ \bibnamefont
  {Kelly}},\ }\href {\doibase 10.1103/PhysRevLett.113.127201} {\bibfield
  {journal} {\bibinfo  {journal} {Phys. Rev. Lett.}\ }\textbf {\bibinfo
  {volume} {113}},\ \bibinfo {pages} {127201} (\bibinfo {year}
  {2014})}.
  %\BibitemShut {NoStop}%

\bibitem [{\citenamefont {Liao}\ \emph {et~al.}(2017)\citenamefont {Liao},
  \citenamefont {Gauquelin}, \citenamefont {Green}, \citenamefont {Macke},
  \citenamefont {Gonnissen}, \citenamefont {Thomas}, \citenamefont {Zhong},
  \citenamefont {Li}, \citenamefont {Si}, \citenamefont {Van~Aert},
  \citenamefont {Hansmann}, \citenamefont {Held}, \citenamefont {Xia},
  \citenamefont {Verbeeck}, \citenamefont {Van~Tendeloo}, \citenamefont
  {Sawatzky}, \citenamefont {Koster}, \citenamefont {Huijben},\ and\
  \citenamefont {Rijnders}}]{Liao2017}%
  %\BibitemOpen
  \bibfield  {author} {\bibinfo {author} {\bibfnamefont {Z.}~\bibnamefont
  {Liao}}, \bibinfo {author} {\bibfnamefont {N.}~\bibnamefont {Gauquelin}},
  \bibinfo {author} {\bibfnamefont {R.~J.}\ \bibnamefont {Green}}, \bibinfo
  {author} {\bibfnamefont {S.}~\bibnamefont {Macke}}, \bibinfo {author}
  {\bibfnamefont {J.}~\bibnamefont {Gonnissen}}, \bibinfo {author}
  {\bibfnamefont {S.}~\bibnamefont {Thomas}}, \bibinfo {author} {\bibfnamefont
  {Z.}~\bibnamefont {Zhong}}, \bibinfo {author} {\bibfnamefont
  {L.}~\bibnamefont {Li}}, \bibinfo {author} {\bibfnamefont {L.}~\bibnamefont
  {Si}}, \bibinfo {author} {\bibfnamefont {S.}~\bibnamefont {Van~Aert}},
  \bibinfo {author} {\bibfnamefont {P.}~\bibnamefont {Hansmann}}, \bibinfo
  {author} {\bibfnamefont {K.}~\bibnamefont {Held}}, \bibinfo {author}
  {\bibfnamefont {J.}~\bibnamefont {Xia}}, \bibinfo {author} {\bibfnamefont
  {J.}~\bibnamefont {Verbeeck}}, \bibinfo {author} {\bibfnamefont
  {G.}~\bibnamefont {Van~Tendeloo}}, \bibinfo {author} {\bibfnamefont {G.~A.}\
  \bibnamefont {Sawatzky}}, \bibinfo {author} {\bibfnamefont {G.}~\bibnamefont
  {Koster}}, \bibinfo {author} {\bibfnamefont {M.}~\bibnamefont {Huijben}}, \
  and\ \bibinfo {author} {\bibfnamefont {G.}~\bibnamefont {Rijnders}},\
  }\href@noop {} {\  (\bibinfo {year} {2017})}.
  %\BibitemShut {NoStop}%

\bibitem [{\citenamefont {Knight}(2011)}]{Knight2011}%
  %\BibitemOpen
  \bibfield  {author} {\bibinfo {author} {\bibfnamefont {K.~S.}\ \bibnamefont
  {Knight}},\ }\href {\doibase 10.1016/j.jallcom.2011.03.014} {\bibfield
  {journal} {\bibinfo  {journal} {J. Alloys Compd.}\ }\textbf {\bibinfo
  {volume} {509}},\ \bibinfo {pages} {6337} (\bibinfo {year}
  {2011})}.
  %\BibitemShut {NoStop}%

\bibitem [{\citenamefont {Ueda}\ \emph {et~al.}(1999)\citenamefont {Ueda},
  \citenamefont {Yanagi}, \citenamefont {Hosono},\ and\ \citenamefont
  {Kawazoe}}]{Ueda1999}%
  %\BibitemOpen
  \bibfield  {author} {\bibinfo {author} {\bibfnamefont {K.}~\bibnamefont
  {Ueda}}, \bibinfo {author} {\bibfnamefont {H.}~\bibnamefont {Yanagi}},
  \bibinfo {author} {\bibfnamefont {H.}~\bibnamefont {Hosono}}, \ and\ \bibinfo
  {author} {\bibfnamefont {H.}~\bibnamefont {Kawazoe}},\ }\href
  {http://iopscience.iop.org/0953-8984/11/17/311} {\bibfield  {journal}
  {\bibinfo  {journal} {J. Phys.: Condensed Matter}\ }\textbf {\bibinfo {volume} {3535}}
  (\bibinfo {year} {1999})}.
  %\BibitemShut {NoStop}%

\bibitem [{\citenamefont {Zhong}\ and\ \citenamefont
  {Kelly}(2008)}]{Zhong2008d}%
  %\BibitemOpen
  \bibfield  {author} {\bibinfo {author} {\bibfnamefont {Z.}~\bibnamefont
  {Zhong}}\ and\ \bibinfo {author} {\bibfnamefont {P.~J.}\ \bibnamefont
  {Kelly}},\ }\href@noop {} {\bibfield  {journal} {\bibinfo  {journal}
  {Europhys. Lett.}\ }\textbf {\bibinfo {volume} {84}},\ \bibinfo {pages}
  {27001} (\bibinfo {year} {2008})}.
  %\BibitemShut {NoStop}%

\bibitem [{\citenamefont {Jia}\ \emph {et~al.}(2009)\citenamefont {Jia},
  \citenamefont {Mi}, \citenamefont {Faley}, \citenamefont {Poppe},
  \citenamefont {Schubert},\ and\ \citenamefont {Urban}}]{Jia2009}%
  %\BibitemOpen
  \bibfield  {author} {\bibinfo {author} {\bibfnamefont {C.~L.}\ \bibnamefont
  {Jia}}, \bibinfo {author} {\bibfnamefont {S.~B.}\ \bibnamefont {Mi}},
  \bibinfo {author} {\bibfnamefont {M.}~\bibnamefont {Faley}}, \bibinfo
  {author} {\bibfnamefont {U.}~\bibnamefont {Poppe}}, \bibinfo {author}
  {\bibfnamefont {J.}~\bibnamefont {Schubert}}, \ and\ \bibinfo {author}
  {\bibfnamefont {K.}~\bibnamefont {Urban}},\ }\href {\doibase
  10.1103/PhysRevB.79.081405} {\bibfield  {journal} {\bibinfo  {journal} {Phys.
  Rev. B}\ }\textbf {\bibinfo {volume} {79}},\ \bibinfo {pages} {081405}
  (\bibinfo {year} {2009})}.
  %\BibitemShut {NoStop}%

\bibitem [{\citenamefont {Rubano}\ \emph {et~al.}(2013)\citenamefont {Rubano},
  \citenamefont {Aruta}, \citenamefont {Uccio}, \citenamefont {Granozio},
  \citenamefont {Marrucci}, \citenamefont {G{\"{u}}nter}, \citenamefont {Fink},
  \citenamefont {Fiebig},\ and\ \citenamefont {Paparo}}]{Rubano2013}%
  %\BibitemOpen
  \bibfield  {author} {\bibinfo {author} {\bibfnamefont {A.}~\bibnamefont
  {Rubano}}, \bibinfo {author} {\bibfnamefont {C.}~\bibnamefont {Aruta}},
  \bibinfo {author} {\bibfnamefont {U.~S.~D.}\ \bibnamefont {Uccio}}, \bibinfo
  {author} {\bibfnamefont {F.~M.}\ \bibnamefont {Granozio}}, \bibinfo {author}
  {\bibfnamefont {L.}~\bibnamefont {Marrucci}}, \bibinfo {author}
  {\bibfnamefont {T.}~\bibnamefont {G{\"{u}}nter}}, \bibinfo {author}
  {\bibfnamefont {T.}~\bibnamefont {Fink}}, \bibinfo {author} {\bibfnamefont
  {M.}~\bibnamefont {Fiebig}}, \ and\ \bibinfo {author} {\bibfnamefont
  {D.}~\bibnamefont {Paparo}},\ }\href {\doibase 10.1103/PhysRevB.88.245434}
  {\bibfield  {journal} {\bibinfo  {journal} {Phys. Rev. B}\ }\textbf {\bibinfo
  {volume} {88}},\ \bibinfo {pages} {245434} (\bibinfo {year}
  {2013})}.
  %\BibitemShut {NoStop}%

\bibitem [{\citenamefont {Chen}\ \emph {et~al.}(2013)\citenamefont {Chen},
  \citenamefont {Luo}, \citenamefont {Ou}, \citenamefont {Yuan}, \citenamefont
  {Wang}, \citenamefont {Yang}, \citenamefont {Yin},\ and\ \citenamefont
  {Liu}}]{Chen2013b}%
  %\BibitemOpen
  \bibfield  {author} {\bibinfo {author} {\bibfnamefont {J.}~\bibnamefont
  {Chen}}, \bibinfo {author} {\bibfnamefont {Y.}~\bibnamefont {Luo}}, \bibinfo
  {author} {\bibfnamefont {X.}~\bibnamefont {Ou}}, \bibinfo {author}
  {\bibfnamefont {G.}~\bibnamefont {Yuan}}, \bibinfo {author} {\bibfnamefont
  {Y.}~\bibnamefont {Wang}}, \bibinfo {author} {\bibfnamefont {Y.}~\bibnamefont
  {Yang}}, \bibinfo {author} {\bibfnamefont {J.}~\bibnamefont {Yin}}, \ and\
  \bibinfo {author} {\bibfnamefont {Z.}~\bibnamefont {Liu}},\ }\href {\doibase
  10.1063/1.4807794} {\bibfield  {journal} {\bibinfo  {journal} {J. Appl.
  Phys.}\ }\textbf {\bibinfo {volume} {113}},\ \bibinfo {pages} {204105}
  (\bibinfo {year} {2013})}.
  %\BibitemShut {NoStop}%

\bibitem [{\citenamefont {Lu}\ \emph {et~al.}(2012)\citenamefont {Lu},
  \citenamefont {Bark}, \citenamefont {{Esque de los Ojos}}, \citenamefont
  {Alcala}, \citenamefont {Eom}, \citenamefont {Catalan},\ and\ \citenamefont
  {Gruverman}}]{Lu2012e}%
  %\BibitemOpen
  \bibfield  {author} {\bibinfo {author} {\bibfnamefont {H.}~\bibnamefont
  {Lu}}, \bibinfo {author} {\bibfnamefont {C.-W.}\ \bibnamefont {Bark}},
  \bibinfo {author} {\bibfnamefont {D.}~\bibnamefont {{Esque de los Ojos}}},
  \bibinfo {author} {\bibfnamefont {J.}~\bibnamefont {Alcala}}, \bibinfo
  {author} {\bibfnamefont {C.~B.}\ \bibnamefont {Eom}}, \bibinfo {author}
  {\bibfnamefont {G.}~\bibnamefont {Catalan}}, \ and\ \bibinfo {author}
  {\bibfnamefont {A.}~\bibnamefont {Gruverman}},\ }\href {\doibase
  10.1126/science.1218693} {\bibfield  {journal} {\bibinfo  {journal} {Science
  (80-. ).}\ }\textbf {\bibinfo {volume} {336}},\ \bibinfo {pages} {59}
  (\bibinfo {year} {2012})}.
  %\BibitemShut {NoStop}%

\bibitem [{\citenamefont {Joshua}\ \emph {et~al.}(2012)\citenamefont {Joshua},
  \citenamefont {Pecker}, \citenamefont {Ruhman}, \citenamefont {Altman},\ and\
  \citenamefont {Ilani}}]{Joshua2012}%
  %\BibitemOpen
  \bibfield  {author} {\bibinfo {author} {\bibfnamefont {A.}~\bibnamefont
  {Joshua}}, \bibinfo {author} {\bibfnamefont {S.}~\bibnamefont {Pecker}},
  \bibinfo {author} {\bibfnamefont {J.}~\bibnamefont {Ruhman}}, \bibinfo
  {author} {\bibfnamefont {E.}~\bibnamefont {Altman}}, \ and\ \bibinfo {author}
  {\bibfnamefont {S.}~\bibnamefont {Ilani}},\ }\href@noop {} {\bibfield
  {journal} {\bibinfo  {journal} {Nat. Commun.}\ }\textbf {\bibinfo {volume}
  {3}},\ \bibinfo {pages} {1129} (\bibinfo {year} {2012})}.
  %\BibitemShut{NoStop}%

\bibitem [{\citenamefont {Joshua}\ \emph {et~al.}(2013)\citenamefont {Joshua},
  \citenamefont {Ruhman}, \citenamefont {Pecker}, \citenamefont {Altman},\ and\
  \citenamefont {Ilani}}]{Joshua2013}%
  %\BibitemOpen
  \bibfield  {author} {\bibinfo {author} {\bibfnamefont {A.}~\bibnamefont
  {Joshua}}, \bibinfo {author} {\bibfnamefont {J.}~\bibnamefont {Ruhman}},
  \bibinfo {author} {\bibfnamefont {S.}~\bibnamefont {Pecker}}, \bibinfo
  {author} {\bibfnamefont {E.}~\bibnamefont {Altman}}, \ and\ \bibinfo {author}
  {\bibfnamefont {S.}~\bibnamefont {Ilani}},\ }\href@noop {} {\bibfield
  {journal} {\bibinfo  {journal} {Proc. Natl. Acad. Sci.}\ }\textbf {\bibinfo
  {volume} {110}},\ \bibinfo {pages} {9633} (\bibinfo {year}
  {2013})}.
  %\BibitemShut {NoStop}%

\bibitem [{\citenamefont {Liang}\ \emph {et~al.}(2015)\citenamefont {Liang},
  \citenamefont {Cheng}, \citenamefont {Wei}, \citenamefont {Luo},
  \citenamefont {Yu}, \citenamefont {Zeng},\ and\ \citenamefont
  {Zhang}}]{Liang2015}%
  %\BibitemOpen
  \bibfield  {author} {\bibinfo {author} {\bibfnamefont {H.}~\bibnamefont
  {Liang}}, \bibinfo {author} {\bibfnamefont {L.}~\bibnamefont {Cheng}},
  \bibinfo {author} {\bibfnamefont {L.}~\bibnamefont {Wei}}, \bibinfo {author}
  {\bibfnamefont {Z.}~\bibnamefont {Luo}}, \bibinfo {author} {\bibfnamefont
  {G.}~\bibnamefont {Yu}}, \bibinfo {author} {\bibfnamefont {C.}~\bibnamefont
  {Zeng}}, \ and\ \bibinfo {author} {\bibfnamefont {Z.}~\bibnamefont {Zhang}},\
  }\href {\doibase 10.1103/PhysRevB.92.075309} {\bibfield  {journal} {\bibinfo
  {journal} {Phys. Rev. B}\ }\textbf {\bibinfo {volume} {92}},\ \bibinfo
  {pages} {075309} (\bibinfo {year} {2015})}.
  %\BibitemShut {NoStop}%

\bibitem [{\citenamefont {Trier}\ \emph {et~al.}(2016)\citenamefont {Trier},
  \citenamefont {Prawiroatmodjo}, \citenamefont {Zhong}, \citenamefont
  {Christensen}, \citenamefont {von Soosten}, \citenamefont {Bhowmik},
  \citenamefont {{Garca Lastra}}, \citenamefont {Chen}, \citenamefont
  {Jespersen},\ and\ \citenamefont {Pryds}}]{Trier2016}%
  %\BibitemOpen
  \bibfield  {author} {\bibinfo {author} {\bibfnamefont {F.}~\bibnamefont
  {Trier}}, \bibinfo {author} {\bibfnamefont {G.~E. D.~K.}\ \bibnamefont
  {Prawiroatmodjo}}, \bibinfo {author} {\bibfnamefont {Z.}~\bibnamefont
  {Zhong}}, \bibinfo {author} {\bibfnamefont {D.~V.}\ \bibnamefont
  {Christensen}}, \bibinfo {author} {\bibfnamefont {M.}~\bibnamefont {von
  Soosten}}, \bibinfo {author} {\bibfnamefont {A.}~\bibnamefont {Bhowmik}},
  \bibinfo {author} {\bibfnamefont {J.~M.}\ \bibnamefont {{Garca Lastra}}},
  \bibinfo {author} {\bibfnamefont {Y.}~\bibnamefont {Chen}}, \bibinfo {author}
  {\bibfnamefont {T.~S.}\ \bibnamefont {Jespersen}}, \ and\ \bibinfo {author}
  {\bibfnamefont {N.}~\bibnamefont {Pryds}},\ }\href@noop {} {\bibfield
  {journal} {\bibinfo  {journal} {arXiv Prepr.}\ ,\ \bibinfo {pages}
  {1603.02850v1}} (\bibinfo {year} {2016})},\ \Eprint
  {http://arxiv.org/abs/arXiv:1603.02850v1} {arXiv:arXiv:1603.02850v1}.
  %\BibitemShut {NoStop}%

\bibitem [{\citenamefont {Herranz}\ \emph {et~al.}(2015)\citenamefont
  {Herranz}, \citenamefont {Singh}, \citenamefont {Bergeal}, \citenamefont
  {Jouan}, \citenamefont {Lesueur}, \citenamefont {G{\'{a}}zquez},
  \citenamefont {Varela}, \citenamefont {Scigaj}, \citenamefont {Dix},
  \citenamefont {S{\'{a}}nchez},\ and\ \citenamefont
  {Fontcuberta}}]{Herranz2015}%
  %\BibitemOpen
  \bibfield  {author} {\bibinfo {author} {\bibfnamefont {G.}~\bibnamefont
  {Herranz}}, \bibinfo {author} {\bibfnamefont {G.}~\bibnamefont {Singh}},
  \bibinfo {author} {\bibfnamefont {N.}~\bibnamefont {Bergeal}}, \bibinfo
  {author} {\bibfnamefont {A.}~\bibnamefont {Jouan}}, \bibinfo {author}
  {\bibfnamefont {J.}~\bibnamefont {Lesueur}}, \bibinfo {author} {\bibfnamefont
  {J.}~\bibnamefont {G{\'{a}}zquez}}, \bibinfo {author} {\bibfnamefont
  {M.}~\bibnamefont {Varela}}, \bibinfo {author} {\bibfnamefont
  {M.}~\bibnamefont {Scigaj}}, \bibinfo {author} {\bibfnamefont
  {N.}~\bibnamefont {Dix}}, \bibinfo {author} {\bibfnamefont {F.}~\bibnamefont
  {S{\'{a}}nchez}}, \ and\ \bibinfo {author} {\bibfnamefont {J.}~\bibnamefont
  {Fontcuberta}},\ }\href {\doibase 10.1038/ncomms7028} {\bibfield  {journal}
  {\bibinfo  {journal} {Nat Commun}\ }\textbf {\bibinfo {volume} {6}},\
  \bibinfo {pages} {6028} (\bibinfo {year} {2015})}.
  %\BibitemShut {NoStop}%
  
%%%%%%%%%%%%%%%%%%%%%%%%% REFERENCES S/M ONLY

\bibitem{Kawasaki1996}%
  %\BibitemOpen
  \bibfield  {author} {\bibinfo {author} {\bibfnamefont {M.}~\bibnamefont
  {Kawasaki}}, \bibinfo {author} {\bibfnamefont {A.}~\bibnamefont {Ohtomo}}, \
  and\ \bibinfo {author} {\bibfnamefont {T.}~\bibnamefont {Arakane}},\ }\href
  {http://www.sciencedirect.com/science/article/pii/S0169433296005120}
  {\bibfield  {journal} {\bibinfo  {journal} {Appl. Surf. Sci.}\ }\textbf
  {\bibinfo {volume} {107}},\ \bibinfo {pages} {102} (\bibinfo {year}
  {1996})}.
  %\BibitemShut {NoStop}%

\bibitem{Walker2015}%
  %\BibitemOpen
  \bibfield  {author} {\bibinfo {author} {\bibfnamefont {S.~M.}\ \bibnamefont
  {Walker}}, \bibinfo {author} {\bibfnamefont {F.~Y.}\ \bibnamefont {Bruno}},
  \bibinfo {author} {\bibfnamefont {Z.}~\bibnamefont {Wang}}, \bibinfo {author}
  {\bibfnamefont {A.}~\bibnamefont {de~la Torre}}, \bibinfo {author}
  {\bibfnamefont {S.}~\bibnamefont {Ricc{\'{o}}}}, \bibinfo {author}
  {\bibfnamefont {A.}~\bibnamefont {Tamai}}, \bibinfo {author} {\bibfnamefont
  {T.~K.}\ \bibnamefont {Kim}}, \bibinfo {author} {\bibfnamefont
  {M.}~\bibnamefont {Hoesch}}, \bibinfo {author} {\bibfnamefont
  {M.}~\bibnamefont {Shi}}, \bibinfo {author} {\bibfnamefont {M.~S.}\
  \bibnamefont {Bahramy}}, \bibinfo {author} {\bibfnamefont {P.~D.~C.}\
  \bibnamefont {King}}, \ and\ \bibinfo {author} {\bibfnamefont
  {F.}~\bibnamefont {Baumberger}},\ }\href {\doibase 10.1002/adma.201501556}
  {\bibfield  {journal} {\bibinfo  {journal} {Adv. Mater.}\ }\textbf {\bibinfo
  {volume} {27}},\ \bibinfo {pages} {3894} (\bibinfo {year}
  {2015})}.
  %\BibitemShut {NoStop}%

%\bibitem{Rodel2016}%
%  %\BibitemOpen
%  \bibfield  {author} {\bibinfo {author} {\bibfnamefont {T.~C.}\ \bibnamefont
%  {R{\"{o}}del}}, \bibinfo {author} {\bibfnamefont {F.}~\bibnamefont
%  {Fortuna}}, \bibinfo {author} {\bibfnamefont {S.}~\bibnamefont {Sengupta}},
%  \bibinfo {author} {\bibfnamefont {E.}~\bibnamefont {Frantzeskakis}}, \bibinfo
%  {author} {\bibfnamefont {P.}~\bibnamefont {{Le F{\`{e}}vre}}}, \bibinfo
%  {author} {\bibfnamefont {F.}~\bibnamefont {Bertran}}, \bibinfo {author}
%  {\bibfnamefont {B.}~\bibnamefont {Mercey}}, \bibinfo {author} {\bibfnamefont
%  {S.}~\bibnamefont {Matzen}}, \bibinfo {author} {\bibfnamefont
%  {G.}~\bibnamefont {Agnus}}, \bibinfo {author} {\bibfnamefont
%  {T.}~\bibnamefont {Maroutian}}, \bibinfo {author} {\bibfnamefont
%  {P.}~\bibnamefont {Lecoeur}}, \ and\ \bibinfo {author} {\bibfnamefont
%  {A.~F.}\ \bibnamefont {Santander-Syro}},\ }\href {\doibase
%  10.1002/adma.201505021} {\bibfield  {journal} {\bibinfo  {journal} {Adv.
%  Mater.}\ }\textbf {\bibinfo {volume} {28}},\ \bibinfo {pages} {1976}
%  (\bibinfo {year} {2016})}.
%  %\BibitemShut {NoStop}%

\bibitem{Jeschke2014}%
  %\BibitemOpen
  \bibfield  {author} {\bibinfo {author} {\bibfnamefont {H.~O.}\ \bibnamefont
  {Jeschke}}, \bibinfo {author} {\bibfnamefont {J.}~\bibnamefont {Shen}}, \
  and\ \bibinfo {author} {\bibfnamefont {R.}~\bibnamefont {Valenti}},\ }\href
  {http://iopscience.iop.org/1367-2630/17/2/023034} {\bibfield  {journal}
  {\bibinfo  {journal} {New J. Phys.}\ }\textbf {\bibinfo {volume} {17}},\
  \bibinfo {pages} {023034} (\bibinfo {year} {2015})},\ \Eprint
  {http://arxiv.org/abs/arXiv:1407.7060v1} {arXiv:arXiv:1407.7060v1}.
  %\BibitemShut {NoStop}%

%\bibitem{Roedel2015}%
%  %\BibitemOpen
%  \bibfield  {author} {\bibinfo {author} {\bibfnamefont {T.~C.}\ \bibnamefont
%  {R{\"{o}}del}}, \bibinfo {author} {\bibfnamefont {F.}~\bibnamefont
%  {Fortuna}}, \bibinfo {author} {\bibfnamefont {F.}~\bibnamefont {Bertran}},
%  \bibinfo {author} {\bibfnamefont {M.}~\bibnamefont {Gabay}}, \bibinfo
%  {author} {\bibfnamefont {M.~J.}\ \bibnamefont {Rozenberg}}, \bibinfo {author}
%  {\bibfnamefont {a.~F.}\ \bibnamefont {Santander-Syro}}, \ and\ \bibinfo
%  {author} {\bibfnamefont {P.~L.}\ \bibnamefont {F{\`{e}}vre}},\ }\href
%  {\doibase 10.1103/PhysRevB.92.041106} {\bibfield  {journal} {\bibinfo
%  {journal} {Phys. Rev. B}\ }\textbf {\bibinfo {volume} {92}},\ \bibinfo
%  {pages} {041106(R)} (\bibinfo {year} {2015})},\ \Eprint
%  {http://arxiv.org/abs/1507.03916} {arXiv:1507.03916}. 
%  %\BibitemShut {NoStop}%

\bibitem{Lechermann2016}%
  %\BibitemOpen
  \bibfield  {author} {\bibinfo {author} {\bibfnamefont {F.}~\bibnamefont
  {Lechermann}}, \bibinfo {author} {\bibfnamefont {H.~O.}\ \bibnamefont
  {Jeschke}}, \bibinfo {author} {\bibfnamefont {A.~J.}\ \bibnamefont {Kim}},
  \bibinfo {author} {\bibfnamefont {S.}~\bibnamefont {Backes}}, \ and\ \bibinfo
  {author} {\bibfnamefont {R.}~\bibnamefont {Valenti}},\ }\href {\doibase
  10.1103/PhysRevB.93.121103} {\bibfield  {journal} {\bibinfo  {journal} {Phys.
  Rev. B}\ }\textbf {\bibinfo {volume} {93}},\ \bibinfo {pages} {121103(R)}
  (\bibinfo {year} {2016})}.
  %\BibitemShut {NoStop}%

\bibitem{Kresse1996}%
  %\BibitemOpen
  \bibfield  {author} {\bibinfo {author} {\bibfnamefont {G.}~\bibnamefont
  {Kresse}}\ and\ \bibinfo {author} {\bibfnamefont {J.}~\bibnamefont
  {Furthmuller}},\ }\href@noop {} {\bibfield  {journal} {\bibinfo  {journal}
  {Phys. Rev. B}\ }\textbf {\bibinfo {volume} {54}},\ \bibinfo {pages} {11169}
  (\bibinfo {year} {1996})}.
  %\BibitemShut {NoStop}%

\bibitem{Kresse1993}%
  %\BibitemOpen
  \bibfield  {author} {\bibinfo {author} {\bibfnamefont {G.}~\bibnamefont
  {Kresse}}\ and\ \bibinfo {author} {\bibfnamefont {J.}~\bibnamefont
  {Hafner}},\ }\href@noop {} {\bibfield  {journal} {\bibinfo  {journal} {Phys.
  Rev. B}\ }\textbf {\bibinfo {volume} {47}},\ \bibinfo {pages} {558} (\bibinfo
  {year} {1993})}.
  %\BibitemShut {NoStop}%

\bibitem{pw91}%
  %\BibitemOpen
  \bibfield  {author} {\bibinfo {author} {\bibfnamefont {J.~P.}\ \bibnamefont
  {Perdew}}, \bibinfo {author} {\bibfnamefont {J.~A.}\ \bibnamefont {Chevary}},
  \bibinfo {author} {\bibfnamefont {S.~H.}\ \bibnamefont {Vosko}}, \bibinfo
  {author} {\bibfnamefont {K.~A.}\ \bibnamefont {Jackson}}, \bibinfo {author}
  {\bibfnamefont {M.~R.}\ \bibnamefont {Pederson}}, \bibinfo {author}
  {\bibfnamefont {D.~J.}\ \bibnamefont {Singh}}, \ and\ \bibinfo {author}
  {\bibfnamefont {C.}~\bibnamefont {Fiolhais}},\ }\href@noop {} {\bibfield
  {journal} {\bibinfo  {journal} {Phys. Rev. B}\ }\textbf {\bibinfo {volume}
  {46}},\ \bibinfo {pages} {6671} (\bibinfo {year} {1992})}.
  %\BibitemShut{NoStop}%

\bibitem{DudarevPRB1998}%
  %\BibitemOpen
  \bibfield  {author} {\bibinfo {author} {\bibfnamefont {S.~L.}\ \bibnamefont
  {Dudarev}}, \bibinfo {author} {\bibfnamefont {G.~A.}\ \bibnamefont {Botton}},
  \bibinfo {author} {\bibfnamefont {S.~Y.}\ \bibnamefont {Savrasov}}, \bibinfo
  {author} {\bibfnamefont {C.~J.}\ \bibnamefont {Humphreys}}, \ and\ \bibinfo
  {author} {\bibfnamefont {A.~P.}\ \bibnamefont {Sutton}},\ }\href@noop {}
  {\bibfield  {journal} {\bibinfo  {journal} {Phys. Rev. B}\ }\textbf {\bibinfo
  {volume} {57}},\ \bibinfo {pages} {1505} (\bibinfo {year}
  {1998})}.
  %\BibitemShut {NoStop}%

\bibitem{hse06a}%
  %\BibitemOpen
  \bibfield  {author} {\bibinfo {author} {\bibfnamefont {J.}~\bibnamefont
  {Heyd}}, \bibinfo {author} {\bibfnamefont {G.~E.}\ \bibnamefont {Scuseria}},
  \ and\ \bibinfo {author} {\bibfnamefont {M.}~\bibnamefont {Ernzerhof}},\
  }\href@noop {} {\bibfield  {journal} {\bibinfo  {journal} {J. Chem. Phys.}\
  }\textbf {\bibinfo {volume} {118}},\ \bibinfo {pages} {8207} (\bibinfo {year}
  {2003})}.
  %\BibitemShut {NoStop}%

\bibitem{hse06b}%
  %\BibitemOpen
  \bibfield  {author} {\bibinfo {author} {\bibfnamefont {J.}~\bibnamefont
  {Paier}}, \bibinfo {author} {\bibfnamefont {M.}~\bibnamefont {Marsman}},
  \bibinfo {author} {\bibfnamefont {K.}~\bibnamefont {Hummer}}, \bibinfo
  {author} {\bibfnamefont {G.}~\bibnamefont {Kresse}}, \bibinfo {author}
  {\bibfnamefont {I.~C.}\ \bibnamefont {Gerber}}, \ and\ \bibinfo {author}
  {\bibfnamefont {J.~G.}\ \bibnamefont {Angyan}},\ }\href@noop {} {\bibfield
  {journal} {\bibinfo  {journal} {J. Chem. Phys.}\ }\textbf {\bibinfo {volume}
  {124}},\ \bibinfo {pages} {154709} (\bibinfo {year} {2006})}.
  % \BibitemShut {NoStop}%

\bibitem{Blochl1994}%
  %\BibitemOpen
  \bibfield  {author} {\bibinfo {author} {\bibfnamefont {P.~E.}\ \bibnamefont
  {Bl{\"{o}}chl}},\ }\href@noop {} {\bibfield  {journal} {\bibinfo  {journal}
  {Phys. Rev. B}\ }\textbf {\bibinfo {volume} {50}},\ \bibinfo {pages} {17953}
  (\bibinfo {year} {1994})}.
  %\BibitemShut {NoStop}%

\bibitem{Kresse1999}%
  %\BibitemOpen
  \bibfield  {author} {\bibinfo {author} {\bibfnamefont {G.}~\bibnamefont
  {Kresse}}\ and\ \bibinfo {author} {\bibfnamefont {J.}~\bibnamefont
  {Joubert}},\ }\href@noop {} {\bibfield  {journal} {\bibinfo  {journal} {Phys.
  Rev. B}\ }\textbf {\bibinfo {volume} {59}},\ \bibinfo {pages} {1758}
  (\bibinfo {year} {1999})}.
  %\BibitemShut {NoStop}%

\bibitem{Glazer1972}%
  %\BibitemOpen
  \bibfield  {author} {\bibinfo {author} {\bibfnamefont {A.~M.}\ \bibnamefont
  {Glazer}},\ }\href@noop {} {\bibfield  {journal} {\bibinfo  {journal} {Acta
  Crystallogr.}\ }\textbf {\bibinfo {volume} {B28}},\ \bibinfo {pages} {3384}
  (\bibinfo {year} {1972})}.
  %\BibitemShut {NoStop}%

\bibitem{Wien2k}%
  %\BibitemOpen
  \bibfield  {author} {\bibinfo {author} {\bibfnamefont {P.}~\bibnamefont
  {Blaha}}, \bibinfo {author} {\bibfnamefont {K.}~\bibnamefont {Schwarz}},
  \bibinfo {author} {\bibfnamefont {G.~K.~H.}\ \bibnamefont {Madsen}}, \bibinfo
  {author} {\bibfnamefont {D.}~\bibnamefont {Kvasnicka}}, \ and\ \bibinfo
  {author} {\bibfnamefont {J.}~\bibnamefont {Luitz}},\ }\href
  {http://www.wien2k.at} {\emph {\bibinfo {title} {{WIEN2k, An Augmented Plane
  Wave and Local Orbitals Program for Calculating Crystal Properties}}}}\
  (\bibinfo  {publisher} {Techn. Universit{{\"a}}t Wien, Austria},\ \bibinfo
  {year} {2001}).
  %\BibitemShut {NoStop}%

\bibitem{Plumb2014}%
  %\BibitemOpen
  \bibfield  {author} {\bibinfo {author} {\bibfnamefont {N.~C.}\ \bibnamefont
  {Plumb}}, \bibinfo {author} {\bibfnamefont {M.}~\bibnamefont {Salluzzo}},
  \bibinfo {author} {\bibfnamefont {E.}~\bibnamefont {Razzoli}}, \bibinfo
  {author} {\bibfnamefont {M.}~\bibnamefont {M{\aa}nsson}}, \bibinfo {author}
  {\bibfnamefont {M.}~\bibnamefont {Falub}}, \bibinfo {author} {\bibfnamefont
  {J.}~\bibnamefont {Krempasky}}, \bibinfo {author} {\bibfnamefont {C.~E.}\
  \bibnamefont {Matt}}, \bibinfo {author} {\bibfnamefont {J.}~\bibnamefont
  {Chang}}, \bibinfo {author} {\bibfnamefont {M.}~\bibnamefont {Schulte}},
  \bibinfo {author} {\bibfnamefont {J.}~\bibnamefont {Braun}}, \bibinfo
  {author} {\bibfnamefont {H.}~\bibnamefont {Ebert}}, \bibinfo {author}
  {\bibfnamefont {J.}~\bibnamefont {Min{\'{a}}r}}, \bibinfo {author}
  {\bibfnamefont {B.}~\bibnamefont {Delley}}, \bibinfo {author} {\bibfnamefont
  {K.-J.}\ \bibnamefont {Zhou}}, \bibinfo {author} {\bibfnamefont
  {T.}~\bibnamefont {Schmitt}}, \bibinfo {author} {\bibfnamefont
  {M.}~\bibnamefont {Shi}}, \bibinfo {author} {\bibfnamefont {J.}~\bibnamefont
  {Mesot}}, \bibinfo {author} {\bibfnamefont {L.}~\bibnamefont {Patthey}}, \
  and\ \bibinfo {author} {\bibfnamefont {M.}~\bibnamefont {Radovi{\'{c}}}},\
  }\href {\doibase 10.1103/PhysRevLett.113.086801} {\bibfield  {journal}
  {\bibinfo  {journal} {Phys. Rev. Lett.}\ }\textbf {\bibinfo {volume} {113}},\
  \bibinfo {pages} {086801} (\bibinfo {year} {2014})}.
  %\BibitemShut {NoStop}%

%\bibitem{Santander-Syro2011}%
%  %\BibitemOpen
%  \bibfield  {author} {\bibinfo {author} {\bibfnamefont {A.~F.}\ \bibnamefont
%  {Santander-Syro}}, \bibinfo {author} {\bibfnamefont {O.}~\bibnamefont
%  {Copie}}, \bibinfo {author} {\bibfnamefont {T.}~\bibnamefont {Kondo}},
%  \bibinfo {author} {\bibfnamefont {F.}~\bibnamefont {Fortuna}}, \bibinfo
%  {author} {\bibfnamefont {S.}~\bibnamefont {Pailh{\`{e}}s}}, \bibinfo {author}
%  {\bibfnamefont {R.}~\bibnamefont {Weht}}, \bibinfo {author} {\bibfnamefont
%  {X.~G.}\ \bibnamefont {Qiu}}, \bibinfo {author} {\bibfnamefont
%  {F.}~\bibnamefont {Bertran}}, \bibinfo {author} {\bibfnamefont
%  {A.}~\bibnamefont {Nicolaou}}, \bibinfo {author} {\bibfnamefont
%  {A.}~\bibnamefont {Taleb-Ibrahimi}}, \bibinfo {author} {\bibfnamefont
%  {P.}~\bibnamefont {{Le F{\`{e}}vre}}}, \bibinfo {author} {\bibfnamefont
%  {G.}~\bibnamefont {Herranz}}, \bibinfo {author} {\bibfnamefont
%  {M.}~\bibnamefont {Bibes}}, \bibinfo {author} {\bibfnamefont
%  {N.}~\bibnamefont {Reyren}}, \bibinfo {author} {\bibfnamefont
%  {Y.}~\bibnamefont {Apertet}}, \bibinfo {author} {\bibfnamefont
%  {P.}~\bibnamefont {Lecoeur}}, \bibinfo {author} {\bibfnamefont
%  {A.}~\bibnamefont {Barth{\'{e}}l{\'{e}}my}}, \ and\ \bibinfo {author}
%  {\bibfnamefont {M.~J.}\ \bibnamefont {Rozenberg}},\ }\href {\doibase
%  10.1038/nature09720} {\bibfield  {journal} {\bibinfo  {journal} {Nature}\
%  }\textbf {\bibinfo {volume} {469}},\ \bibinfo {pages} {189} (\bibinfo {year}
%  {2011})}.
%  %\BibitemShut {NoStop}%
  
\end{thebibliography}

%
%%%%%%%%%%%%%

\end{document}